\documentclass[aps,pre,onecolumn,floatfix]{revtex4}

\usepackage{graphicx}
\usepackage{amssymb,amsfonts,amsmath}
\usepackage{epsf}
\usepackage{subfigure}
\usepackage{epstopdf}

\newcommand{\be}{\begin{equation}}
\newcommand{\ee}{\end{equation}}
\newcommand{\bea}{\begin{eqnarray}}
\newcommand{\eea}{\end{eqnarray}}

\begin{document}

\title{\bf Growth and dissolution of macromolecular Markov chains}

\author{Pierre Gaspard}
\affiliation{Center for Nonlinear Phenomena and Complex Systems,\\
Universit\'e Libre de Bruxelles, Code Postal 231, Campus Plaine,
B-1050 Brussels, Belgium}

\begin{abstract}
The kinetics and thermodynamics of free living copolymerization are studied for processes with rates depending on $k$ monomeric units of the macromolecular chain behind the unit that is attached or detached.  In this case, the sequence of monomeric units in the growing copolymer is a $k^{\rm th}$-order Markov chain.  In the regime of steady growth, the statistical properties of the sequence are determined analytically in terms of the attachment and detachment rates.  In this way, the mean growth velocity as well as the thermodynamic entropy production and the sequence disorder can be calculated systematically.  These different properties are also investigated in the regime of depolymerization where the macromolecular chain is dissolved by the surrounding solution.  In this regime, the entropy production is shown to satisfy Landauer's principle.
\end{abstract}

\noindent 
\vskip 0.5 cm

\maketitle

\section{Introduction}

The growth of macromolecular chains composed of different species of monomeric units is a kinetic process that is fundamental in polymer science and biology \cite{F53,O04,ABLRRW89}.  Such macromolecules are called copolymers or heteropolymers, as opposed to homopolymers composed of identical monomeric units \cite{RMJB85,JKSS96}.  In general, the sequence of monomeric units is irregular in copolymers and a key issue is to understand their statistical properties.  They can be obtained experimentally by sequencing methods or NMR analysis, revealing that the sequences follow statistical laws described as Markov chains of different orders \cite{R78,RR94,SM04}.  In principle, these statistical properties are determined by the conditions prevailing to the synthesis of the copolymers.  

The complexity of such kinetic processes holds in the fact that they generate a multitude of possible sequences.  These processes are thus defined in the infinite space of all the corresponding concentrations, which all change in time.  Typically, the kinetic equations are nonlinear if the reaction network is such that the sequences may attach between each other (and reversely break into any possible subsequences), which is favored by chain transfer reactions.  In the case of living copolymerization, the reaction proceeds without chain termination, without chain transfer reaction, and after a rapid chain initiation avoiding nucleation \cite{JKSS96}.  This is the case if the process evolves only by the attachment and detachment of monomeric units to the chain, for instance with a catalyst located at the tip of the copolymer.  Free living copolymerization is most important in polymer science where macromolecules composed of various types of monomeric units can be synthesized with newly advanced organometallic catalyst systems \cite{OCY97,YMIMSMMNTOYMF04,ZN05,CNS06,HLLLL09,LZFWLZ13,MSBB13,PAP09,DK09}.

Since the forties, the kinetic theory of living copolymerization has been developed for the fully irreversible regime where the chain grows only by the attachment of monomeric units, ignoring the detachment events \cite{ML44,AG44,FR99}.  Remarkably, the kinetics can also be solved analytically in the presence of both detachment and attachment events, as recently shown for the growth of Bernoulli and first-order Markov chains \cite{AG09,GA14}.

The purpose of the present paper is to extend these results to living copolymerization with attachment and detachment rates depending not only on the last monomeric unit incorporated in the chain, but also on $k$ previously incorporated monomeric units.  As suggested by previous work~\cite{AG09,GA14} devoted to the $k=0$ and $k=1$ cases, $k^{\rm th}$-order Markov chains are generated if the kinetics depends on $k$ lastly incorporated monomeric units \cite{SSOL15}.  In the present paper, the analytical method previously developed in Ref.~\cite{GA14} is generalized from first to arbitrary order $k$.  The central idea of this method is that the properties of the growing chain can be deduced by solving a set of self-consistent equations, directly established in terms of the attachment and detachment rates \cite{GA14}.  This theoretical method is much more efficient than numerical Monte Carlo simulations.  In this way, the composition of the macromolecular chain can be determined, as well as the mean growth velocity and the corresponding thermodynamic quantities, including the entropy production.  A fluctuation relation is also established.

Furthermore, the dissolution of the macromolecular chain is investigated and analytical formulas are obtained for the mean depolymerization velocity and the thermodynamic entropy production, generalizing previous results as well \cite{GA14,AG13}.  The erasing of information possibly contained in the macromolecular sequence is shown to obey Landauer's principle \cite{L61}, and also a stronger bound coming from the order $k$ of the reaction network.  There is an interesting analogy with the growth and dissolution of crystals, which are the three-dimensional extensions of the one-dimensional macromolecular chains that are here studied.  

In the following developments, a more abstract analogy is made between the sequences of monomeric units in the macromolecular chains and the symbolic dynamics of chaotic systems \cite{N95}, the analogy providing useful techniques for mathematical analysis.

The paper is organized as follows.  Section~\ref{Growth} is devoted to the growth regime.  In Subsection~\ref{Growth-Kinetics} and Appendix~\ref{AppA}, the kinetic equations are solved in the regime of steady growth.  In Subsection~\ref{Growth-Thermo}, the results are used to obtain the thermodynamic entropy production,  necessary and sufficient conditions for equilibrium, and a fluctuation relation that is partly deduced in Appendix~\ref{AppB}.  Section~\ref{Dissolution} is dealing with the dissolution of the macromolecular chains by depolymerization.  The kinetics of this regime is solved in Subsection~\ref{Dissolution-Kinetics} and the thermodynamics is obtained in Subsection~\ref{Dissolution-Thermo}.  Section~\ref{Examples} compares the theoretical results with kinetic Monte Carlo simulations for two illustrative examples.  Section~\ref{Conclusion} contains the conclusion.

\section{Growth of $k^{\rm th}$-order Markov chains}
\label{Growth}

\subsection{Kinetics}
\label{Growth-Kinetics}

\subsubsection{Master equations}

The growth proceeds by living copolymerization, i.e., the successive attachments and detachments of different species of monomers $m_j\in\{1,2,...,M\}$ to macromolecular chains $\omega=m_1m_2\cdots m_l$, elongating without termination.  The chains -- also called copolymers -- and the monomers are in solution where they move and react together.  The growth is assumed to take place only at one of both ends of each chain:
\be
\omega=m_1m_2 \cdots m_{l-1} \ + \ m_l 
\quad \rightleftharpoons \quad 
\omega'=m_1m_2 \cdots m_{l-1}m_l \, .
\label{reactions}
\ee
The rates of attachment and detachment, respectively $w_{\pm m_l\vert m_{l-1}\cdots m_{l-k}}$, depend not only on the monomer $m_l$ that is on the verge of being attached or detached, but also on $k$ previously incorporated monomeric units $m_{l-k}\cdots m_{l-1}$.  The number~$k$ determines the order of the Markov chain ruling the copolymer sequence, as will be shown in detail here below using the method of Ref.~\cite{GA14}.  The number of monomeric units in the macromolecular chain $l=\vert\omega\vert\in{\mathbb N}$ defines the length of the copolymer. The chain is growing if attachment is more frequent than detachment, otherwise the chain dissolves by depolymerization.  The reactions~(\ref{reactions}) may be induced by a catalyst located at the growing tip of the copolymer.

The process is supposed to be isothermal at the temperature $T$ and to evolve in a dilute solution.  If attachment and detachment are the elementary steps of the reaction~(\ref{reactions}), the rates are determined by the mass action law.  In this respect, the attachment rates are proportional to the concentrations $c_m$ of the monomers in solution, while the detachment rates are independent of the concentrations.  Moreover, if an external force $f$ is exerted on the tip of the copolymer or the catalyst therein located, the rates have an Arrhenius-type dependence on the change of activation energy induced by the external force, which is often exponential \cite{KF07}.  Under such circumstances, the rates can be modeled by
\bea
\mbox{attachment rates:}\qquad &&w_{+m_l\vert m_{l-1}\cdots m_{l-k}} = k_{+m_l\vert m_{l-1}\cdots m_{l-k}}  \, \exp\left(\beta\, f \, d_{+m_l\vert m_{l-1}\cdots m_{l-k}}\right) \, c_{m_l} \, ,\label{w+}\\
\mbox{detachment rates:}\qquad &&w_{-m_l\vert m_{l-1}\cdots m_{l-k}} = k_{-m_l\vert m_{l-1}\cdots m_{l-k}} \,  \exp\left(-\beta\, f \, d_{-m_l\vert m_{l-1}\cdots m_{l-k}}\right) ,\label{w-}
\eea
where $k_{\pm m_l\vert m_{l-1}\cdots m_{l-k}}$ are the rate constants, $\beta=T^{-1}$ the inverse temperature in units where Boltzmann's constant is equal to unity,  $d_{\pm m_l\vert m_{l-1}\cdots m_{l-k}}$ the transition-state displacements in the direction of the external force $f$, and $c_{m_l}$ the concentration of the monomer species that is attached \cite{KF07,G15NJP}.  The solution is large enough so that the concentrations $c_m$ of the monomers remain constant during the whole process.  If the reactions~(\ref{reactions}) are not elementary, the attachment and detachment rates are more complicated than in Eqs.~(\ref{w+}) and~(\ref{w-}), for instance, with Michaelis-Menten dependences on the monomer concentrations.  However, the following considerations remain general because they hold without specifying the dependences of the rates $w_{\pm m_l\vert m_{l-1}\cdots m_{l-k}}$, except otherwise stated.

In the solution, the copolymers are dilute enough so that they do not interact with each other and the process can be considered at the level of a single copolymer.  In this case, the concentration of copolymers with the length $l$ and the sequence $\omega=m_1m_2\cdots m_l $ is proportional to the probability that a single copolymer has these properties at the time $t$:
\be
c_t(m_1m_2\cdots m_l ) = \frac{N}{V} \, P_t(m_1 \cdots m_l) \, ,
\label{conc}
\ee
where $N$ is the total number of copolymers in the solution of volume $V$.
This probability evolves in time according to the master equations:
\bea
\frac{d}{dt}\, P_t(m_1\cdots m_{l-1}m_l) &=& w_{+m_l\vert m_{l-1}\cdots m_{l-k}} \, P_t(m_1\cdots m_{l-1}) \nonumber\\
&& +\sum_{m_{l+1}=1}^M w_{-m_{l+1}\vert m_{l}\cdots m_{l-k+1}} \, P_t(m_1\cdots m_{l-1}m_{l}m_{l+1}) \nonumber\\
&& -\left( w_{-m_{l}\vert m_{l-1}\cdots m_{l-k}} + \sum_{m_{l+1}=1}^M w_{+m_{l+1}\vert m_{l}\cdots m_{l-k+1}}\right)  P_t(m_1\cdots m_{l-1}m_{l}) \, , \nonumber\\
&&
\label{master}
\eea
forming an infinite hierarchy of coupled linear ordinary differential equations of first order in time.  Accordingly, the growth is described as a stochastic jump process on the coarse-grained states:
\be
\omega\in\{\emptyset, \, m_1, \, m_1m_2, \, m_1m_2m_3,\ldots ,\, m_1m_2\cdots m_l,\ldots\} \, ,
\label{seq}
\ee
which form a tree growing as $M^l$ with the length $l$.
Equation~(\ref{master}) holds for $l=k+1,k+2,\ldots.$  For $l=0,1,2,\ldots,k$, the same expression can still be used with $m_0=\emptyset$, $m_{-1}=\emptyset$, ... , $m_{-k}=\emptyset$, ... by defining the following initiation rates $w_{\pm m_k\vert m_{k-1}\cdots m_{2}m_{1}\emptyset}$, $w_{\pm m_{k-1}\vert m_{k-2}\cdots m_{1}\emptyset\emptyset}$, $w_{\pm m_{1}\vert \emptyset\cdots \emptyset\emptyset\emptyset}$, and by setting $w_{\pm \emptyset\vert \emptyset\cdots \emptyset\emptyset\emptyset}=0$.  Consequently, the total probability is conserved, $\sum_\omega P_t(\omega)=1$, with the sum running over the states~(\ref{seq}).

In order to describe the process at the growing tip of the copolymer, the hierarchy of master equations~(\ref{master}) can be transformed into another hierarchy for the probabilities:
\bea
p_t(l) &\equiv& \sum_{m_1\cdots m_l} P_t(m_1\cdots m_l) \, , \label{p0} \\
p_t(m_l,l) &\equiv& \sum_{m_1\cdots m_{l-1}} P_t(m_1\cdots m_l) \, , \label{p1}\\
p_t(m_{l-1}m_l,l) &\equiv& \sum_{m_1\cdots m_{l-2}} P_t(m_1\cdots m_l) \, , \label{p2} \\
&\vdots& \nonumber
\eea
They are interpreted as follows:  $p_t(l)$ is the probability that the copolymer has the length $l$ at the time $t$; $p_t(m_l,l)$ the probability that it has the length $l$ and $m_l$ is the last monomeric unit at its tip;
$p_t(m_{l-1}m_l,l)$ is the probability that it has the length $l$, its last monomeric unit is $m_l$, and its penultimate unit is $m_{l-1}$; etc.  These probabilities are related to each other according to
\bea
p_t(l) &=& \sum_{m_l} p_t(m_l,l) \, , \label{p0-p1} \\
p_t(m_l,l) &=& \sum_{m_{l-1}} p_t(m_{l-1}m_l,l) \, , \label{p1-p2}\\
p_t(m_{l-1}m_l,l) &=& \sum_{m_{l-2}} p_t(m_{l-2}m_{l-1}m_l,l) \, , \label{p2-p3}\\
&\vdots& \nonumber
\eea
Regrouping these probabilities into the infinite vector
\be
{\bf P}_t(l) = \left\{ p_t(l), \, p_t(m_l,l), \, p_t(m_{l-1}m_l,l),\, \ldots\, \right\}  \, ,
\ee
the full set of equations can be written in the form:
\be
\frac{d}{dt}\, {\bf P}_t(l)  = {\boldsymbol{\mathsf A}}\cdot{\bf P}_t(l-1) +{\boldsymbol{\mathsf B}}\cdot{\bf P}_t(l+1) -{\boldsymbol{\mathsf C}}\cdot{\bf P}_t(l) 
\label{Eq-P}
\ee
with infinite matrices ${\boldsymbol{\mathsf A}}$, ${\boldsymbol{\mathsf B}}$, and ${\boldsymbol{\mathsf C}}$ of constant coefficients given by the rates.  

\subsubsection{Solutions of the master equations}

Since Eq.~(\ref{Eq-P}) is linear in the probabilities, its general solution can be decomposed as a linear superposition of the particular solutions:
\be
{\bf P}_t(l) = \exp(s_q t + iql) \, \pmb{\Phi}_q 
\label{part_sol_1}
\ee
expressed in terms of the wavenumber $-\pi < q \leq +\pi$ and a dispersion relation, which is expected to take the form:
\be
s_q = - i \, v \, q  - {\cal D}\, q^2 + O(q^3)  \, ,
\ee
because the growth process is similar to a random walk of drift velocity $v$ and diffusivity $\cal D$ to be determined.  The drift velocity $v$ is the mean growth velocity or mean elongation rate counted in monomeric units incorporated in the macromolecular chain per unit time.   Equation~(\ref{part_sol_1}) can also be expressed as
\be
p_t(m_{l-r+1}\cdots m_{l-1}m_l,l) =  \exp(s_q t + iql) \, \phi_q(m_{l-r+1}\cdots m_{l-1}m_l)
\label{part_sol_2}
\ee
for the probability to find the subsequence $m_{l-r+1}\cdots m_{l-1}m_l$ at the growing tip of the copolymer
with $r=0,1,2,3,...$.  We notice that the quantities $\phi_q(m_{l-r+1}\cdots m_{l-1}m_l)$ are related to each other by Eqs.~(\ref{p0-p1}),~(\ref{p1-p2}),~(\ref{p2-p3}),... in the same way as the corresponding probabilities are.  Inserting the particular solution~(\ref{part_sol_1}) into Eq.~(\ref{Eq-P}), the dispersion relation is given by the eigenvalue problem:
\be
\left( {\boldsymbol{\mathsf A}}\, {\rm e}^{-iq} +{\boldsymbol{\mathsf B}}\, {\rm e}^{+iq} -{\boldsymbol{\mathsf C}}\right)\cdot \pmb{\Phi}_q  = s_q \, \pmb{\Phi}_q \, .
\label{Eq-P-eigen}
\ee
The right and left eigenvectors are respectively denoted $\pmb{\Phi}_q$ and $\tilde{\pmb{\Phi}}_q$.
By expanding these eigenvectors and the eigenvalue~$s_q$ in powers of the wavenumber $q$ and taking the limit $q=0$, Eq.~(\ref{Eq-P-eigen}) gives the mean growth velocity as
\be
v = \frac{\tilde{\pmb{\Phi}}_0^{\rm T}\cdot ({\boldsymbol{\mathsf A}}-{\boldsymbol{\mathsf B}})\cdot\pmb{\Phi}_0}{\tilde{\pmb{\Phi}}_0^{\rm T}\cdot\pmb{\Phi}_0}  \, ,
\ee
where
\bea
 ({\boldsymbol{\mathsf A}}+{\boldsymbol{\mathsf B}}-{\boldsymbol{\mathsf C}})\cdot\pmb{\Phi}_0&=&0  \, ,\label{Phi_0}\\
\tilde{\pmb{\Phi}}_0^{\rm T}\cdot ({\boldsymbol{\mathsf A}}+{\boldsymbol{\mathsf B}}-{\boldsymbol{\mathsf C}})&=&0  \, .
\eea
Because of probability conservation, the left eigenvector at $q=0$ has identical elements, which can be fixed to the unit value: $\tilde{\pmb{\Phi}}_0=\{1,1,1,...\}$.  Renormalizing the right eigenvector at $q=0$, the following probabilities are defined
\bea
\mu(m_{l}) &\equiv& \phi_0(m_{l})/\phi_0(\emptyset) \, , \label{mu1} \\
\mu(m_{l-1}m_{l}) &\equiv& \phi_0(m_{l-1}m_{l})/\phi_0(\emptyset) \, , \label{mu2}\\
\mu(m_{l-2}m_{l-1}m_{l}) &\equiv& \phi_0(m_{l-2}m_{l-1}m_{l})/\phi_0(\emptyset) \, , \label{mu3}\\
&\vdots& \nonumber
\eea
which satisfy
\bea
\sum_{m_l}\mu(m_l) &=& 1  \, ,\\
\sum_{m_{l-1}}\mu(m_{l-1}m_l) &=& \mu(m_l)  \, ,\label{mu-mu}\\
\sum_{m_{l-2}}\mu(m_{l-2}m_{l-1}m_l) &=& \mu(m_{l-1}m_l)  \, ,\label{mu-mu-mu}\\
&\vdots&  \nonumber
\eea
as the consequence of Eqs.~(\ref{p0-p1}),~(\ref{p1-p2}),~(\ref{p2-p3}), etc.  These quantities have the following interpretations: In the regime of steady growth,  $\mu(m_l)$ is the probability to find the monomeric unit $m_l$ at the tip of the copolymer, $\mu(m_{l-1}m_l)$ is the probability that $m_l$ is the last monomeric unit and $m_{l-1}$ is the penultimate unit of the chain, etc. 

Supposing that these time-independent probabilities are known, a closed equation can be deduced for the time-dependent probability $p_t(l)$ that the chain has the length $l\gg k$ at the time $t$:
\be
\frac{d}{dt} p_t(l) = a \, p_t(l-1) + b\, p_t(l+1) - (a+b) \, p_t(l) \, ,
\label{Eq-p(l)}
\ee
where the coefficients are constant and expressed as
\bea
a &=& \sum_{m_l m_{l-1}\cdots m_{l-k}} w_{+m_l\vert m_{l-1}\cdots m_{l-k}} \, \mu(m_{l-k}\cdots m_{l-1}) \, , \\
b &=& \sum_{m_l m_{l-1}\cdots m_{l-k}} w_{-m_l\vert m_{l-1}\cdots m_{l-k}} \, \mu(m_{l-k}\cdots m_{l-1}m_{l}) \, .
\eea
Solving Eq.~(\ref{Eq-p(l)}), the dispersion relation is obtained as
\be
s_q = (a+b)\, (\cos q-1)- i \,(a-b) \,\sin q 
\label{dispersion2}
\ee
with $-\pi < q \leq +\pi$, whereupon the mean growth velocity is given by
\bea
v &=& a-b \nonumber\\
&=& \sum_{m_l m_{l-1}\cdots m_{l-k}} w_{+m_l\vert m_{l-1}\cdots m_{l-k}} \, \mu(m_{l-k}\cdots m_{l-1}) \nonumber\\
&& -\sum_{m_l m_{l-1}\cdots m_{l-k}} w_{-m_l\vert m_{l-1}\cdots m_{l-k}} \, \mu(m_{l-k}\cdots m_{l-1}m_{l})
\label{v}
\eea
and the diffusivity by ${\cal D}=(a+b)/2$.  As a consequence, the length probability distribution obeys the central limit theorem:
\be
{\rm Prob}\left(\frac{l-vt}{\sqrt{2{\cal D}t}}<x\right) = \sum_{l=0}^{vt+x\sqrt{2{\cal D}t}} p_t(l) \longrightarrow_{t\to\infty} \frac{1}{\sqrt{2\pi}} \int_{-\infty}^x \exp(-y^2/2) \, dy \, ,
\label{CLT}
\ee
meaning that, in the long-time limit, the random variable $x\equiv (l-vt)/\sqrt{2{\cal D}t}$ has a Gaussian distribution of zero mean and unit variance~\cite{F68}.  Accordingly, the mean length is growing linearly in time as $\langle l\rangle_t\simeq v\, t$, and the standard deviation as
\be
\sqrt{\langle l^2\rangle_t-\langle l\rangle_t^2} \simeq \sqrt{2{\cal D}t} \, ,
\label{width}
\ee
where the notation $\langle l^n\rangle_t\equiv \sum_l  p_t(l)\, l^n$ is used with $n=1,2$.
We notice that, beyond the time scale $t\simeq k^2/(2{\cal D})$, the width~(\ref{width}) of the length probability distribution becomes larger than the number~$k$ of monomeric units determining the rates, which justifies the steady growth at a constant mean velocity $v$ after a long enough time.  In this regime, the probabilities can therefore be supposed to behave as
\be
P_t(m_1\cdots m_{l-1}m_l) \simeq p_t(l)\, \mu(m_1\cdots m_{l-1}m_l)
\label{p-mu_factor}
\ee
in terms of the time-dependent length probability distribution~(\ref{p0}) and the time-independent probabilities~(\ref{mu1}), (\ref{mu2}), (\ref{mu3}), etc, which is commonly assumed \cite{CF63JPS,CF63JCP}.

Now, these stationary probabilities are obtained by solving the hierarchy of Eqs.~(\ref{Phi_0}), which are explicitly written down in Appendix~\ref{AppA} for the probabilities $\mu(m_{l-r+1}\cdots m_{l-1}m_l)$ of subsequences with an increasing number $r$ of monomeric units.  The key observation is that these equations are similar to each other for $r=k+1,k+2,...$, which shows that they can be solved if  the growing copolymer is assumed to be a $k^{\rm th}$-order Markov chain, implying the factorization:
\bea
&&\mu(m_1m_2m_3\cdots m_{l-2}m_{l-1}m_l) =\nonumber\\ &&\qquad\qquad\mu(m_1\vert m_2 \cdots m_{k+1}) \, \mu(m_2\vert m_3 \cdots m_{k+2}) \cdots \mu(m_{l-k}\vert m_{l-k+1} \cdots m_{l}) \, \mu(m_{l-k+1} \cdots m_{l})
\nonumber\\
&&\label{Markov-factor}
\eea
in terms of the $M^{k+1}$ conditional probabilities $\mu(m_{l-k}\vert m_{l-k+1} \cdots m_{l})$
and the $M^k$ tip probabilities $\mu(m_{l-k+1} \cdots m_{l})$.  These probabilities are normalized according to
\bea
&&\sum_{m_{l-k}} \mu(m_{l-k}\vert m_{l-k+1} \cdots m_{l}) = 1 \, ,
\label{cond_prob_norm} \\
&&\sum_{m_{l-k+1}\cdots m_l} \mu(m_{l-k+1} \cdots m_{l}) = 1 \, .
\label{tip_prob_norm}
\eea

Using the factorization~(\ref{Markov-factor}) in the second term of the expression~(\ref{v}) for the mean growth velocity, this latter can be written as
\be
v = \sum_{m_{l-k}\cdots m_{l-1}} v_{m_{l-k}\cdots m_{l-1}} \, \mu(m_{l-k}\cdots m_{l-1})
\label{v2}
\ee
by defining the $M^k$ {\it partial velocities}:
\be
v_{m_{l-k}\cdots m_{l-1}} \equiv \sum_{m_l} w_{+m_l\vert m_{l-1}\cdots m_{l-k}}  -  \sum_{m_l} w_{-m_l\vert m_{l-1}\cdots m_{l-k}} \, \mu(m_{l-k}\vert m_{l-k+1}\cdots m_{l}) \, \frac{\mu(m_{l-k+1}\cdots m_{l})}{\mu(m_{l-k}\cdots m_{l-1})} \, .
\label{part_v}
\ee
As shown in Appendix~\ref{AppA}, these partial velocities are directly determined in terms of the attachment and detachment rates by solving the $M^k$ following self-consistent equations:
\be
\boxed{
v_{m_{l-k}\cdots m_{l-1}} = \sum_{m_l} \frac{w_{+m_l\vert m_{l-1}\cdots m_{l-k}}\, v_{m_{l-k+1}\cdots m_{l}}}{w_{-m_l\vert m_{l-1}\cdots m_{l-k}}+v_{m_{l-k+1}\cdots m_{l}}}
}
\label{iter}
\ee
which play a central role in the following developments.
In the growth regime, these equations can be solved by numerical iteration in the $M^k$-dimensional space of positive partial velocities, which shows convergence.  Once the partial velocities are known, the tip probabilities are obtained by solving the $M^k$ following equations:
\be
\mu(m_{l-k+1} \cdots m_{l}) = \sum_{m_{l-k}} \frac{w_{+m_l\vert m_{l-1}\cdots m_{l-k}}}{w_{-m_l\vert m_{l-1}\cdots m_{l-k}}+v_{m_{l-k+1}\cdots m_{l}}} \, \mu(m_{l-k}\cdots m_{l-1}) \, ,
\label{tip_prob}
\ee
and the $M^{k+1}$ conditional probabilities are given by
\be
\mu(m_{l-k}\vert m_{l-k+1} \cdots m_{l}) = \frac{w_{+m_l\vert m_{l-1}\cdots m_{l-k}}\, \mu(m_{l-k}\cdots m_{l-1})}{(w_{-m_l\vert m_{l-1}\cdots m_{l-k}}+v_{m_{l-k+1}\cdots m_{l}})\, \mu(m_{l-k+1}\cdots m_{l})} \, ,
\label{cond_prob}
\ee
which completely determines the $k^{\rm th}$-order Markov chain.  The mean growth velocity is finally obtained with Eq.~(\ref{v2}) by averaging the partial velocities over the tip probabilities~(\ref{tip_prob}).

We notice that Eqs.~(\ref{tip_prob}) are recovered by multiplying Eq.~(\ref{cond_prob}) with the tip probability $\mu(m_{l-k+1}\cdots m_{l})$ of the denominator, summing over $m_{l-k}$, and using the normalization~(\ref{cond_prob_norm}) of the conditional probabilities.

The bulk probabilities are defined as the stationary probabilities of the Markov chain according to
\be
\bar\mu(m_{j} \cdots m_{j+k-1}) = \sum_{m_{j+k}} \mu(m_{j}\vert m_{j+1} \cdots m_{j+k})  \, \bar\mu(m_{j+1} \cdots m_{j+k}) \, ,
\label{chain_bulk_prob}
\ee
which gives the probability to find the subsequence $m_{j} \cdots m_{j+k-1}$ of length $k$, anywhere in a long enough chain of length $l\gg k$.  The bulk and tip probabilities are related to each other by
\be
\bar\mu(m_{l-k+1} \cdots m_{l}) = \frac{1}{v} \, v_{m_{l-k+1}\cdots m_{l}}  \, \mu(m_{l-k+1} \cdots m_{l}) \, .
\label{bulk_prob}
\ee
Indeed, replacing Eq.~(\ref{bulk_prob}) into Eq.~(\ref{chain_bulk_prob}) and using Eq.~(\ref{cond_prob}), we go back to the self-consistent equations~(\ref{iter}).  Moreover, the expression~(\ref{v2}) for the mean growth velocity is recovered by summing Eq.~(\ref{bulk_prob}) over $m_{l-k+1} \cdots m_{l}$, which completes the proof of Eq.~(\ref{bulk_prob}).

\subsubsection{Correlation functions characterizing the sequences}

We notice that the statistical properties of the copolymer can be characterized in terms of correlation functions:
\be
C_{AB}(j) =  \left\langle [ A(\omega) -\langle A\rangle ] [B(\sigma^j\omega) -\langle B\rangle]\right\rangle
\label{correl-AB}
\ee
between some observables $A(\omega)$ and $B(\omega)$ defined over the subsequences $\omega=m_i\cdots m_{i+k-1}$.  In Eq.~(\ref{correl-AB}), $\sigma$ denotes the shift of the subsequence by one monomeric unit $\sigma\omega=m_{i+1}\cdots m_{i+k}$.  The decay of such correlation functions is determined by the eigenvalues $\{\Lambda_{\alpha}\}$ of the $M^k\times M^k$ matrix
\be
{\boldsymbol{\mathsf M}} = \left\{\mu(m_{l-k}\vert m_{l-k+1}\cdots m_{l})\, \delta_{m_{l-k}\cdots m_{l-1},m_{l-k+1}\cdots m_l} \right\} \, ,
\label{M}
\ee
formed by the conditional probabilities~(\ref{cond_prob}).  We notice that the leading eigenvalue is equal to unity, $\Lambda_1=1$, due to the conservation of probability~(\ref{cond_prob_norm}).

These results have already been established for first-order Markov chains in the case $k=1$ \cite{GA14}.

\subsubsection{The fully irreversible growth regime}

In this regime, the detachment of the last monomeric unit becomes negligible with respect to the attachment of new units so that the detachment rates can be assumed to be vanishing:
\be
w_{-m_l\vert m_{l-1}\cdots m_{l-k}} = 0 \, .
\label{w-_irr}
\ee
Hence, the partial velocities are simply given by summing the attachment rates according to Eq.~(\ref{part_v}):
\be
v_{m_{l-k}\cdots m_{l-1}} = \sum_{m_l} w_{+m_l\vert m_{l-1}\cdots m_{l-k}} \, .
\label{part_v_irr}
\ee
The corresponding tip and conditional probabilities are inferred by Eqs.~(\ref{tip_prob})-(\ref{cond_prob}) under the conditions~(\ref{w-_irr}).

In this regime, which has been much studied since the forties, we recover previously established results, in particular, the famous Mayo-Lewis formula for first-order Markov chain \cite{ML44, AG44, FR99}.

\subsection{Thermodynamics}
\label{Growth-Thermo}

\subsubsection{Entropy production}

The entropy production is given by
\bea
\frac{d_{\rm i}S}{dt} &=& \sum_{m_1\cdots m_l} \left[
w_{+m_l\vert m_{l-1}\cdots m_{l-k}} \, P_t(m_1\cdots m_{l-1}) - w_{-m_l\vert m_{l-1}\cdots m_{l-k}} \, P_t(m_1\cdots m_{l-1}m_l)\right] \nonumber\\
&&\qquad\qquad\qquad\qquad \times \ \ln\frac{w_{+m_l\vert m_{l-1}\cdots m_{l-k}} \, P_t(m_1\cdots m_{l-1})}{w_{-m_l\vert m_{l-1}\cdots m_{l-k}} \, P_t(m_1\cdots m_{l-1}m_l)} \geq 0 \, ,
\label{entrprod1}
\eea
which is equivalent to the expression known for dilute solutions in terms of the concentrations~(\ref{conc}) \cite{P55,S76,N79,JQQ04}.  In the long-time limit, the probability distribution factorizes according to Eq.~(\ref{Markov-factor}) and the length probability distribution becomes broader and broader because of the central limit theorem~(\ref{CLT}) and Eq.~(\ref{width}).  Therefore, we may suppose that $p_t(l-1)\simeq p_t(l)$.  Hence, the entropy production reaches the following stationary value
\bea
\frac{d_{\rm i}S}{dt} &=& \sum_{m_{l-k}\cdots m_l} J_{m_{l-k}\cdots m_{l}} \, A_{m_{l-k}\cdots m_{l}} \geq 0 \, ,
\label{entrprod2}
\eea
expressed in terms of the net rates
\be
J_{m_{l-k}\cdots m_{l}} \equiv w_{+m_l\vert m_{l-1}\cdots m_{l-k}} \, \mu(m_{l-k}\cdots m_{l-1}) - w_{-m_l\vert m_{l-1}\cdots m_{l-k}} \, \mu(m_{l-k}\vert m_{l-k+1}\cdots m_l) \, \mu(m_{l-k+1}\cdots m_l)
\label{J}
\ee
and the affinities
\be
A_{m_{l-k}\cdots m_{l}} \equiv \ln\frac{w_{+m_l\vert m_{l-1}\cdots m_{l-k}} \, \mu(m_{l-k}\cdots m_{l-1})}{w_{-m_l\vert m_{l-1}\cdots m_{l-k}} \, \mu(m_{l-k}\vert m_{l-k+1}\cdots m_l) \, \mu(m_{l-k+1}\cdots m_l)} \, .
\label{affinities}
\ee
The net rates (\ref{J}) can be written first in terms of the partial velocities by using Eq.~(\ref{cond_prob}) and, thereafter, in terms of the bulk probabilities with Eq.~(\ref{bulk_prob}) to get
\be
J_{m_{l-k}\cdots m_{l}} = v\,  \bar\mu(m_{l-k+1} \cdots m_{l})\, \mu(m_{l-k}\vert m_{l-k+1}\cdots m_l) \, .
\label{J2}
\ee
Now, the affinities~(\ref{affinities}) are expanded as follows
\be
A_{m_{l-k}\cdots m_{l}} =\ln\frac{w_{+m_l\vert m_{l-1}\cdots m_{l-k}}}{w_{-m_l\vert m_{l-1}\cdots m_{l-k}}} 
- \ln\mu(m_{l-k}\vert m_{l-k+1}\cdots m_l) +\ln\mu(m_{l-k}\cdots m_{l-1}) -\ln\mu(m_{l-k+1}\cdots m_l) \, .
\label{affinities-expanded}
\ee
Replacing into Eq.~(\ref{entrprod2}), we find that the last two terms cancel each other because of Eq.~(\ref{chain_bulk_prob}) and the normalization conditions~(\ref{cond_prob_norm}).  We thus find that the entropy production takes the form
\be
\frac{d_{\rm i}S}{dt} = v \, \left( \epsilon + D \right) \geq 0 \, ,
\label{entrprod3}
\ee
in terms of the mean growth velocity $v$, the free-energy driving force \cite{B79}
\be
\epsilon = -\beta g = \sum_{m_{l-k}\cdots m_l} \bar\mu(m_{l-k+1}\cdots m_{l})\ \mu(m_{l-k}\vert m_{l-k+1} \cdots m_{l}) \ \ln\frac{w_{+m_l\vert m_{l-1}\cdots m_{l-k}}}{w_{-m_l\vert m_{l-1}\cdots m_{l-k}}} \, ,
\label{eps}
\ee
and the Shannon disorder per monomeric unit
\be
D = - \sum_{m_{l-k}\cdots m_l} \bar\mu(m_{l-k+1}\cdots m_{l})\ \mu(m_{l-k}\vert m_{l-k+1} \cdots m_{l}) \ \ln\mu(m_{l-k}\vert m_{l-k+1}\cdots m_l) \geq 0 \, ,
\label{D}
\ee
as established in Ref.~\cite{AG08}.  The free-energy driving force is proportional to the mean free-enthalpy $g$ per incorporated monomeric unit.  Combining Eqs.~(\ref{eps}) and~(\ref{D}) defines the mean affinity:
\be
A = \epsilon+D = \sum_{m_{l-k}\cdots m_l} \bar\mu(m_{l-k+1} \cdots m_{l})\, \mu(m_{l-k}\vert m_{l-k+1}\cdots m_l) \, A_{m_{l-k}\cdots m_{l}} \, ,
\label{A}
\ee
which is the thermodynamic entropy production per incorporated monomeric unit.
We recover the expressions of Refs.~\cite{AG09,GA14} in the cases of the processes with $k=0$ and $k=1$, respectively generating Bernoulli chains and first-order Markov chains.  Since the entropy production~(\ref{entrprod3}) is always non-negative according to the second law of thermodynamics, the free-energy driving force is bounded as
\be
-D \leq \epsilon \, , \qquad\mbox{if}\quad v\geq 0 \, ,
\ee
i.e., in the growth regime.  Therefore, the growth of the copolymer can be driven not only by a positive free-energy driving force $\epsilon >0$, but also by the entropic effect of sequence disorder if $-D < \epsilon \leq 0$ \cite{AG08}.  

The thermodynamic entropy production~(\ref{entrprod3}) becomes infinite in the fully irreversible regime, because the free-energy driving force~(\ref{eps}) diverges under the conditions~(\ref{w-_irr}), although the mean growth velocity and the Shannon disorder keep finite values.

\subsubsection{Equilibrium}

The principle of detailed balance holds at thermodynamic equilibrium so that the net rates~(\ref{J}) and the affinities~(\ref{affinities}) both vanish.  Consequently, the conditional probabilities are given by
\be
\mu_{\rm eq}(m_{l-k}\vert m_{l-k+1}\cdots m_l) = \frac{w_{+m_l\vert m_{l-1}\cdots m_{l-k}} \, \mu_{\rm eq}(m_{l-k}\cdots m_{l-1})}{w_{-m_l\vert m_{l-1}\cdots m_{l-k}} \, \mu_{\rm eq}(m_{l-k+1}\cdots m_l)} \, .
\label{cond_prob_eq}
\ee
Comparing with Eq.~(\ref{cond_prob}), we can conclude that the partial velocities are also vanishing at equilibrium: $v_{m_{l-k+1}\cdots m_{l},{\rm eq}}=0$.  Therefore, the mean growth velocity is vanishing $v=0$ together with the entropy production and the affinity.  As a consequence, the free-energy driving force is related to the Shannon disorder per monomeric unit by $\epsilon_{\rm eq}=-D_{\rm eq}$.

Summing Eq.~(\ref{cond_prob_eq}) over $m_{l-k}$ and using the normalization conditions~(\ref{cond_prob_norm}), we see that the equilibrium tip probability should satisfy
\be
\sum_{m_{l-k}} z_{m_l\vert m_{l-1}\cdots m_{l-k}} \, \mu_{\rm eq}(m_{l-k}\cdots m_{l-1}) = \mu_{\rm eq}(m_{l-k+1}\cdots m_l) 
\label{tip_prob_eq}
\ee
with the coefficients
\be
z_{m_l\vert m_{l-1}\cdots m_{l-k}} \equiv \frac{w_{+m_l\vert m_{l-1}\cdots m_{l-k}}}{w_{-m_l\vert m_{l-1}\cdots m_{l-k}}} \, .
\label{z}
\ee
The following $M^k\times M^k$ matrix is introduced
\be
{\boldsymbol{\mathsf Z}} = \left\{z_{m_l\vert m_{l-1}\cdots m_{l-k}}\, \delta_{m_{l-k}\cdots m_{l-1},m_{l-k+1}\cdots m_l}\right\} ,
\label{MZ}
\ee
which is composed of the elements~(\ref{z}) located at the row $m_{l-k}\cdots m_{l-1}$ and the column $m_{l-k+1}\cdots m_l$.  The necessary condition to solve the set~(\ref{tip_prob_eq}) of homogeneous linear equations is that
\be
\det({\boldsymbol{\mathsf Z}}-{\boldsymbol{\mathsf 1}})=0 \, ,
\label{Z-eq}
\ee
implying that at least one of the eigenvalues $z_\alpha$ of the matrix~(\ref{MZ}) is equal to unity. 

Beyond, a necessary and sufficient condition for the system to be at equilibrium is that the spectral radius of the matrix~(\ref{MZ}) is equal to unity:
\be
\rho({\boldsymbol{\mathsf Z}}) \equiv{\rm max}_{\alpha}\{\vert z_\alpha\vert\} = 1 \, .
\label{NSC_eq}
\ee
The copolymer is growing if $\rho({\boldsymbol{\mathsf Z}})>1$ and depolymerizing if $\rho({\boldsymbol{\mathsf Z}})<1$.

\subsubsection{Fluctuation relation}

A remarkable result is that the incorporation of monomeric units in the chain is ruled by a fluctuation relation, which is established as follows in the framework of large-deviation theory~\cite{E85,T09}.  

Suppose that the chain has been growing in the past till reaching the length $l_0\gg 1$.  From this initial situation, the growth process goes on and we count the $M^{k+1}$ numbers $\Delta{\bf N} = \left\{ \Delta N_{m_j\cdots m_{j+k}}\right\}$ of the different possible subsequences $m_j\cdots m_{j+k}$ of length $k+1$ that are formed at the tip of the chain every time a new monomeric unit is incorporated.  Due to the possible detachment events, these numbers can take negative values if the chain length decreases and some multiplets of length $k+1$ disappear.  Accordingly, the counters may fluctuate in time between positive and negative values $\Delta{\bf N}\in{\mathbb Z}^{M^{k+1}}$, even if they grow on average.  As shown in Appendix~\ref{AppB}, the probability $p_t(\Delta{\bf N})$ that these counters take the values $\Delta{\bf N}$ at the time $t$ obeys the fluctuation relation:
\be
\frac{p_t(\Delta{\bf N})}{p_t(-\Delta{\bf N})}\simeq_{t\to\infty} {\rm e}^{{\bf A}\cdot\Delta{\bf N}} \, ,
\label{FT}
\ee
where ${\bf A}=\left\{A_{m_j\cdots m_{j+k}}\right\}$ denotes the $M^{k+1}$ affinities~(\ref{affinities}).  This result generalizes the relations already obtained in the special cases of Bernoulli and first-order Markov chains to $k^{\rm th}$-order Markov chains with $k>1$ \cite{G15NJP,G15EPJST}. 

We notice that the fluctuation relation~(\ref{FT}) implies the non-negativity of the entropy production in accordance with the second law of thermodynamics.  Indeed, the entropy production can be obtained in terms of the Kullback-Leibler divergence between the probability distributions $p_t(\Delta{\bf N})$ and $p_t(-\Delta{\bf N})$:
\be
\frac{d_{\rm i}S}{dt} =\lim_{t\to\infty} \frac{1}{t} \sum_{\Delta{\bf N}} p_t(\Delta{\bf N}) \, \ln \frac{p_t(\Delta{\bf N})}{p_t(-\Delta{\bf N})} \geq 0 \, .
\label{KL-N}
\ee
Replacing the ratio of probabilities in the logarithm by the fluctuation relation~(\ref{FT}), we find that the entropy production can be expressed as
\be
\frac{d_{\rm i}S}{dt} ={\bf A}\cdot{\bf J} = A \, v \geq 0
\label{entrprod-N}
\ee
in terms of the net rates~(\ref{J}) or~(\ref{J2}) given by 
\be
{\bf J} = \lim_{t\to\infty}\frac{1}{t}\langle\Delta{\bf N}\rangle_t \equiv \lim_{t\to\infty} \frac{1}{t} \sum_{\Delta{\bf N}} p_t(\Delta{\bf N}) \, \Delta{\bf N}\, ,
\ee
which is equivalent to Eq.~(\ref{entrprod3}) in terms of the mean growth velocity~(\ref{v}) and the mean affinity~(\ref{A}).  In this regard, the fluctuation relation~(\ref{FT}) concerns the currents of multiplets incorporated in the chain.

\section{Dissolution of $k^{\rm th}$-order Markov chains}
\label{Dissolution}

The dissolution of macromolecular chains is the process that is reverse to the growth.  This dissolution proceeds by the depolymerization of the copolymer due to the dominance of monomeric detachment  over attachment.  In this section, we generalize the results of Refs.~\cite{GA14,AG13} to depolymerization processes where the attachment and detachment rates $w_{\pm m_l\vert m_{l-1}\cdots m_{l-k}}$ depend  on the last $k+1$ monomeric units $m_{l-k}\cdots m_{l-1}m_l$ at the tip of the copolymer. 

Contrary to the growth where the copolymer is generated by a kinetic process with properties essentially independent of the initial conditions, the depolymerization does depend on the initial copolymer.  Indeed, this latter is introduced in the solution at the beginning of dissolution and it has thus been previously synthesized under conditions different from those of the solution where depolymerization takes place.  Consequently, the rate of depolymerization and the other properties depend on the initial sequence of the copolymer.  This latter is supposed to be long enough so that a regime of steady depolymerization can be reached before the complete dissolution of the initial copolymer.

\subsection{Kinetics}
\label{Dissolution-Kinetics}

The mean depolymerization velocity $-v>0$ can be obtained by a first-passage problem \cite{KT75,KT81}.  For non-vanishing attachment rates, the depolymerization is a succession of detachment and attachment events.  A transient growth may happen between two detachment events.  The mean time before the monomeric unit $m_l$ is detached is given by summing over all the possible transient growths during which $r=0,1,2,3,...$ monomeric units are attached.  If $r=0$, the removal of $m_l$ lasts the time $(w_{-m_l\vert m_{l-1}\cdots m_{l-k}})^{-1}$.  If $r=1$, the transient growth is caused by the transitions $m_l\to m_lm_{l+1}\to m_l$, so that the removal lasts the time $(w_{-m_l\vert m_{l-1}\cdots m_{l-k}})^{-1}\sum_{m_{l+1}} z_{m_{l+1}\vert m_i \cdots m_{l-k+1}}$, which is expressed in terms of the ratios~(\ref{z}) of attachment to detachment rates.  If $r>1$, the transient growth continues up to the elongation of the copolymer by the subsequence $m_l m_{l+1}\cdots m_{l+r}$, so that the corresponding time is given by a similar expression with a product of ratios~(\ref{z}) and a sum over the monomeric units $m_{l+1}\cdots m_{l+r}$.  Adding all these contributions, the mean time is obtained as
\be
\langle T_{m_l\vert m_{l-1}\cdots m_{l-k}}\rangle = \frac{1}{w_{-m_l\vert m_{l-1}\cdots m_{l-k}}} \left(1 +\sum_{r=1}^{\infty} \sum_{m_{l+1}\cdots m_{l+r}} \prod_{i=l}^{l+r-1} z_{m_{i+1}\vert m_i \cdots m_{i-k+1}}\right) .
\label{mean-time}
\ee

During depolymerization, such transient growths may occur for all the successive monomeric units of the initial copolymer.  If $\bar\mu_{k+1}(m_0m_1\cdots m_k)$ denotes the probability to find the multiplet $m_0m_1\cdots m_k$ in the sequence of the initial copolymer, the mean depolymerization velocity is thus given by averaging as
\be
-v = \left[\sum_{m_0m_1\cdots m_k} \bar\mu_{k+1}(m_0m_1\cdots m_k) \, \langle T_{m_k\vert m_{k-1}\cdots m_0}\rangle\right]^{-1}  .
\ee
Using the matrix notation~(\ref{MZ}), we find that
\be
-v = \left[\sum_{m_0m_1\cdots m_k} \frac{\bar\mu_{k+1}(m_0m_1\cdots m_k)}{w_{-m_k\vert m_{k-1}\cdots m_0}} \, \sum_{n_1\cdots n_k} \left(\frac{{\boldsymbol{\mathsf 1}}}{{\boldsymbol{\mathsf 1}}-{\boldsymbol{\mathsf Z}}}\right)_{n_k\cdots n_1, m_k\cdots m_1}\right]^{-1}  ,
\label{v-depolym}
\ee
with $m_j,n_j\in\{1,2,...,M\}$.  The mean velocity $v$ is negative because $\rho({\boldsymbol{\mathsf Z}})<1$ in the depolymerization regime.

\subsection{Thermodynamics}
\label{Dissolution-Thermo}

Since the initial copolymer has a specific sequence, there is no sequence disorder that is generated during depolymerization contrary to what happens during growth.  In the regime of steady depolymerization, the thermodynamic entropy production thus takes the form
\be
\frac{d_{\rm i}S}{dt} = v \, \bar\epsilon \geq 0 \, ,
\label{entrprod-depolym}
\ee
where the free-energy driving force is here given by
\be
\bar\epsilon = -\beta \bar{g} = \sum_{m_0m_1\cdots m_k} \bar\mu_{k+1}(m_0m_1\cdots m_k) \ \ln\frac{w_{+m_k\vert m_{k-1}\cdots m_0}}{w_{-m_k\vert m_{k-1}\cdots m_0}} \, ,
\label{eps-depolym}
\ee
in terms of the probability $\bar\mu_{k+1}(m_0m_1\cdots m_k)$ of the corresponding multiplet in the initial copolymer.

The free-energy driving force~(\ref{eps-depolym}) is bounded by Landauer's principle \cite{L61}, according to which erasing possible information contained in the initial copolymer during its dissolution has a thermodynamic cost.  This bound can be established by considering the Kullback-Leibler divergence (per monomeric unit) between the sequence probability distribution of the initial copolymer and the one of the copolymer that would be grown at thermodynamic equilibrium:
\be
D_{\rm KL}\left[\bar\mu\Vert\mu^{\rm (eq)}\right] = \lim_{l\to\infty} \frac{1}{l} \sum_{\omega} \bar\mu_l(\omega) \, \ln\frac{\bar\mu_l(\omega)}{\mu_{l}^{\rm (eq)}(\omega)} \geq 0 \, .
\label{KL-div-1}
\ee
According to the principle of detailed balance, the equilibrium probability distribution is determined by Eq.~(\ref{cond_prob_eq}) in terms of the ratios of transition rates and it is related to the free enthalpy $G(\omega)$ of the copolymer in the solution by
\bea
\mu_{l}^{\rm (eq)}(\omega) &=& \prod_{j=1}^{l-k} \mu_{\rm eq}(m_{j}\vert m_{j+1}\cdots m_{j+k}) \, \mu_{\rm eq}(m_{l-k+1}\cdots m_l)\nonumber\\
&=&  \mu_{\rm eq}(m_1\cdots m_k) \, \prod_{j=1}^{l-k}  \frac{w_{+m_{j+k}\vert m_{j+k-1}\cdots m_{j}}}{w_{-m_{j+k}\vert m_{j+k-1}\cdots m_{j}}} \nonumber \\
&\sim& \exp\left[-\beta\, G(\omega)\right] \, .
\label{mu_eq}
\eea
The Kullback-Leibler divergence~(\ref{KL-div-1}) thus becomes
\be
D_{\rm KL}\left[\bar\mu\Vert\mu^{\rm (eq)}\right] = \beta\, \bar{g} - \bar{I}_{\infty}\geq 0 \, ,
\label{KL-div-2}
\ee
where $\bar{g}=-\beta^{-1}\bar{\epsilon}$ is the free enthalpy per monomeric unit averaged over the sequence of the initial copolymer and given by~(\ref{eps-depolym}), while
\be
\bar{I}_{\infty} \equiv \lim_{l\to\infty} -\frac{1}{l} \sum_{\omega} \bar\mu_l(\omega) \, \ln\bar\mu_l(\omega) \geq 0 
\label{I-infty}
\ee
is the Shannon disorder per monomeric unit of the initial copolymer.  This quantity also characterizes the information possibly coded in the initial copolymer and that is erased during depolymerization.  Therefore, the general bound implied by Landauer's principle is that the entropy production given by Eq.~(\ref{entrprod-depolym}) should remain larger than the Shannon information, both evaluated per monomeric unit:
\be
\frac{1}{\vert v\vert} \, \frac{d_{\rm i}S}{dt} \geq \bar{I}_{\infty} \geq 0 \, .
\label{Landauer}
\ee

However, the depolymerization process only depends on no more than $k+1$ monomeric units at the copolymer tip.  We can thus also consider the Kullback-Leibler divergence between the sequence probability distribution of the initial copolymer truncated to a $k^{\rm th}$-order Markov chain and the equilibrium one to get
\be
D_{\rm KL}\left[\bar\mu^{(k)}\Vert\mu^{\rm (eq)}\right] = \beta\, \bar{g} - \bar{I}_{k+1}\geq 0 
\label{KL-div-k-3}
\ee
with the Shannon information contained in the multiplets of length $k+1$:
\be
\bar{I}_{k+1} \equiv- \sum_{m_0m_1\cdots m_k} \bar\mu_{k+1}(m_0m_1\cdots m_k) \, \ln\bar\mu_{k+1}(m_0\vert m_1\cdots m_k)\geq 0 \, .
\label{I-k+1}
\ee
The thermodynamic entropy production per monomeric unit is thus bounded by
\be
\frac{1}{\vert v\vert} \, \frac{d_{\rm i}S}{dt} \geq \bar{I}_{k+1} \geq 0 \, ,
\label{bound-k+1}
\ee
which holds if the rates depend on no more than $k+1$ monomeric units behind the tip of the copolymer. 

In general, $\bar{I}_{k+1} \geq \bar{I}_{\infty}$ so that the bound~(\ref{bound-k+1}) is stronger than~(\ref{Landauer}) and Landauer's principle is always obeyed.

\section{Illustrative examples}
\label{Examples}

In this section, several illustrative examples are considered for the growth and depolymerization of $2^{\rm nd}$-order Markov chains and composed of two monomeric species $m\in\{ 1,2\}$, so that $k=2$ and $M=2$.  The previous theoretical results are compared with numerical simulations using Gillespie's kinetic Monte Carlo algorithm \cite{G76,G77}.  In each figure, the dots depict data from numerical simulations and the lines the corresponding theoretical results.  In the illustrative examples, the external force is set equal to zero $f=0$ and the mass action law applies.  The attachment rates are thus proportional to the concentrations of the monomers that are attached, while the detachment rates do not depend on the concentrations according to Eqs.~(\ref{w+})-(\ref{w-}).

\subsection{Growth of period-two copolymers and dissolution}
\label{Subsec-Ex1}

In the first example, the rate constants are taken as
\be
\begin{array}{llll}
k_{+1\vert 11}=0.05 \, , &
k_{+1\vert 12}=0.1 \, , &
k_{+1\vert 21}=2 \, , &
k_{+1\vert 22}=0.5 \, , \\
k_{+2\vert 11}=1  \, , &
k_{+2\vert 12}=3  \, , &
k_{+2\vert 21}=0.4  \, , &
k_{+2\vert 22}=0.1  \, , \\
k_{-1\vert 11}=0.002 \, , &
k_{-1\vert 12}=0.001 \, , &
k_{-1\vert 21}=0.02 \, , &
k_{-1\vert 22}=0.03 \, , \\
k_{-2\vert 11}=0.003 \, , &
k_{-2\vert 12}=0.001 \, , &
k_{-2\vert 21}=0.01 \, , &
k_{-2\vert 22}=0.04 \, .
\end{array}
\label{Ex1}
\ee
The concentration $c_1$ of monomers $1$ is varying, while the concentration of monomers $2$ is fixed to the value: $c_2=0.005$.  The attachment and detachment rates are thus given respectively by $w_{+m_3\vert m_2m_1}=k_{+m_3\vert m_2m_1}\, c_{m_3}$ and $w_{-m_3\vert m_2m_1}=k_{-m_3\vert m_2m_1}$ with $m_1,m_2,m_3\in\{1,2\}$.

\subsubsection{Growth}

The example~(\ref{Ex1}) illustrates the growth of an alternating copolymer, which would be of period two $(12)^{\infty}$ if all the rates were equal to zero except $k_{\pm 1\vert 21}$ and $k_{\pm 2\vert 12}$.  Since the other rates are non-vanishing, the growing chains are not perfectly periodic and some disorder manifests itself.

In the case where $k=2$ and $M=2$, the self-consistent equations~(\ref{iter}) for the partial velocities are given by
\be
v_{mn} = \frac{w_{+1\vert nm}\, v_{n1}}{w_{-1\vert nm}+v_{n1}} + \frac{w_{+2\vert nm}\, v_{n2}}{w_{-2\vert nm}+v_{n2}} \qquad\mbox{for} \quad m,n\in\{1,2\} \, .
\label{iter2}
\ee
These equations are solved by numerical iterations starting from positive initial values.  
Away from equilibrium, convergence up to ten digits is achieved in a few dozens of iterations.  At equilibrium where Eq.~(\ref{NSC_eq}) is satisfied, the convergence is slower and goes as the inverse of the number of iterations.  An accuracy of ten digits is obtained over the whole concentration range with a thousand iterations.

\begin{figure}[h]
\centerline{\scalebox{0.5}{\includegraphics{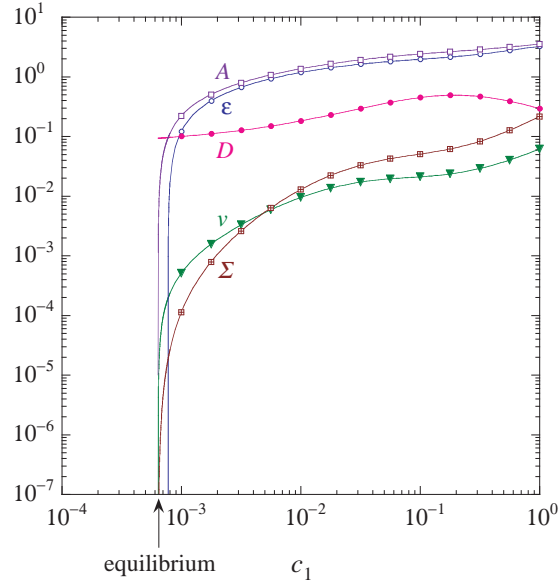}}}
\caption{Growth of a copolymer with the rate constants~(\ref{Ex1}) and the concentration $c_2=0.005$:  The mean growth velocity $v$, the free-energy driving force $\epsilon$, the Shannon disorder per monomeric unit $D$, the mean affinity $A=\epsilon+D$, and the thermodynamic entropy production $\Sigma=d_{\rm i}S/dt=A\, v$ are plotted versus the concentration $c_1$.  The dots are the numerical simulation data obtained with a sample of $10^4$ sequences grown with $10^6$ reactive events.  The solid lines are the corresponding theoretical results.}
\label{fig1}
\end{figure}

Once the partial velocities are obtained, the tip and conditional probabilities are successively calculated with Eqs.~(\ref{tip_prob}) and~(\ref{cond_prob}).  The mean growth velocity is given by Eq.~(\ref{v2}), the free energy driving force by Eq.~(\ref{eps}), the Shannon disorder per monomeric unit by Eq.~(\ref{D}), the mean affinity by the sum $A=\epsilon+D$, and the thermodynamic entropy production by Eq.~(\ref{entrprod3}).  These different quantities are shown in Fig.~\ref{fig1} as a function of the concentration~$c_1$ together with the results of kinetic Monte Carlo simulations with Gillespie's algorithm.  The agreement between theory and the numerical results is excellent.  Using a thousand iterations of Eqs.~(\ref{iter2}), the different quantities of interest can be obtained more than $10^6$ faster and with a higher accuracy than with the Monte Carlo simulation of $10^4$ sequences.  

In Fig.~\ref{fig1}, we see that the mean growth velocity, the mean affinity, and the entropy production are vanishing at the critical concentration $c_{1,{\rm eq}}\simeq 0.00064027$ corresponding to thermodynamic equilibrium.  This critical concentration is calculated by solving Eq.~(\ref{Z-eq}) and selecting the value where the necessary and sufficient condition~(\ref{NSC_eq}) is satisfied.  At equilibrium, the Shannon disorder per monomeric unit takes the value $D_{\rm eq}=-\epsilon_{\rm eq}\simeq 0.09276$.  However, the free-energy driving force is vanishing $\epsilon=0$ at the larger concentration $c_{1,\epsilon=0}\simeq 0.00077711$.  In the interval $c_{1,{\rm eq}} < c_1 \leq c_{1,\epsilon=0}$, the growth is driven by the entropic effect of sequence disorder \cite{AG08}.  The free-energy driving force becomes positive and plays its role for larger concentrations: $c_{1,\epsilon=0} < c_1$.

\begin{figure}[h]
\centerline{\scalebox{0.5}{\includegraphics{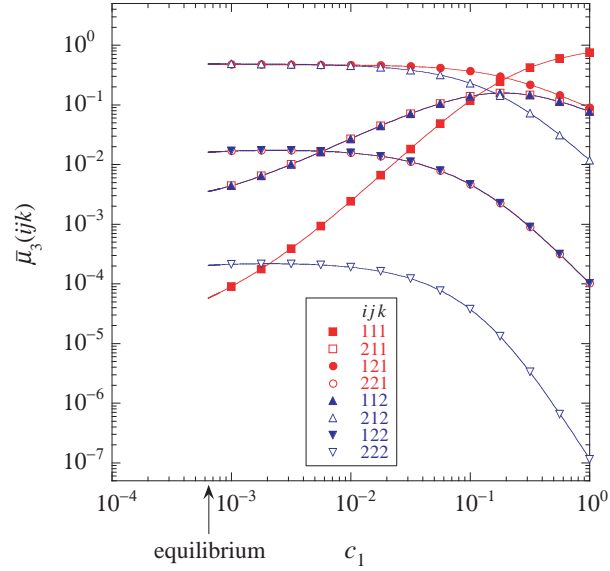}}}
\caption{Growth of a copolymer with the rate constants~(\ref{Ex1}) and the concentration $c_2=0.005$:  The bulk probabilities $\bar\mu_3(m_1m_2m_3)$ of the triplets $ijk=m_1m_2m_3$ are plotted versus the concentration $c_1$.  The dots are the numerical simulation data obtained with a sample of $10^4$ sequences grown with $10^6$ reactive events.  The solid lines are the corresponding theoretical results.}
\label{fig2}
\end{figure}

Figure~\ref{fig2} depicts the bulk probabilities $\bar\mu_3(m_1m_2m_3)$ of the triplets $m_1m_2m_3$, as given by Eq.~(\ref{bulk_prob}).  We observe that the bulk probability $\bar\mu_3(111)$ is the largest  if $c_1>0.2$, under which conditions the attachment of monomers $1$ dominates the growth process.  However, the bulk probabilities $\bar\mu_3(121)$ and $\bar\mu_3(212)$ become the largest for $c_1 <0.2$ and they reach the values $\bar\mu_3(121)\simeq 0.475$ and $\bar\mu_3(212)\simeq 0.486$, nearly sharing the whole probability close to equilibrium.  This is evidence for an alternating copolymer of near periodicity $(12)^{\infty}$.

\begin{figure}[h]
\centerline{\scalebox{0.5}{\includegraphics{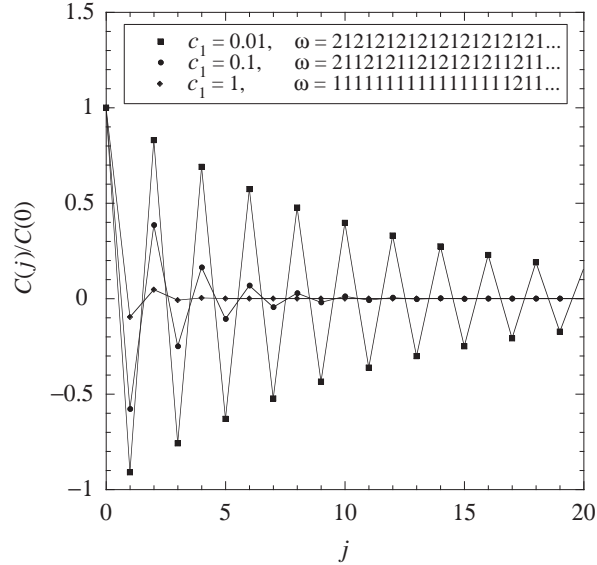}}}
\caption{Growth of a copolymer with the rate constants~(\ref{Ex1}) and the concentration $c_2=0.005$:  The normalized correlation functions~(\ref{correl-num}) versus the distance $j$ between successive monomeric units for the concentrations $c_1=0.01, 0.1,$~and~$1$. The dots are the numerical simulation data obtained with a sample of $N=10^3$ sequences of total length $L\simeq 3.7\times 10^5$ if $c_1=0.01$, $L\simeq 5.8\times 10^5$ if $c_1=0.1$, and $L\simeq 9.2\times 10^5$ if $c_1=1$.  The solid lines are the corresponding theoretical results.  The inset shows an example of the growing sequence $\omega$ for each concentration.}
\label{fig3}
\end{figure}

This is confirmed by the correlation functions
\be
C(j) =  \frac{1}{N} \sum_{n=1}^N \left[ \frac{1}{L}\sum_{i=1}^L m_i^{(n)} m_{i+j}^{(n)} -\left(\frac{1}{L}\sum_{i=1}^L m_i^{(n)}\right)^2 \right] 
\label{correl-num}
\ee
computed with $N$ sequences of total length $L$.  They are shown in Fig.~\ref{fig3} for the concentrations $c_1=0.01, 0.1$ and~$1$.  The alternating character manifests itself by the fact that the correlation function changes sign, which is due to the presence of a negative leading eigenvalue for the matrix~(\ref{M}) formed by the conditional probabilities $\mu(m_1\vert m_2m_3)$.  The calculation of these eigenvalues again confirms this conclusion:
\be
\begin{array}{llllll}
&c_1 = 0.01  : & \Lambda_1=1 \, , & \Lambda_2\simeq -0.912 \, ,& \Lambda_3\simeq 0.028 \, ,& \Lambda_4 \simeq -0.021 \, ; \\
&c_1 = 0.1  : & \Lambda_1=1 \, ,& \Lambda_2\simeq -0.651 \, ,& \Lambda_3\simeq 0.120 \, ,& \Lambda_4 \simeq -0.005 \, ; \\
&c_1 = 1  : & \Lambda_1=1 \, ,& \Lambda_2\simeq -0.248 \, ,& \Lambda_3\simeq 0.155 \, ,& \Lambda_4 \simeq 10^{-6} \, .
\end{array}
\ee
We note that the leading eigenvalue $\Lambda_2$ is significantly smaller in absolute value if the concentration is $c_1=1$, which explains that the corresponding correlation function decays faster in Fig.~\ref{fig3} and the alternating character is much less marked in this case where the monomeric unit $1$ is most frequent in the chain since $\bar\mu_3(111)\simeq 0.745$ when $c_1=1$.

\subsubsection{Dissolution}

For lower concentrations $c_1<c_{1,{\rm eq}}$, the chain undergoes depolymerization.  The initial chain should thus have a long enough length $l_0$, which decreases during the process simulated with the same algorithm as the growth.  In order to show the important dependence on the sequence of the initial copolymer, depolymerization is compared between three different initial sequences: a Bernoulli chain of probabilities $\bar\mu_1(1)=\bar\mu_1(2)=0.5$ with the Shannon informations $\bar I_{\infty}=\bar I_3=\ln 2$; the periodic chain $(11121222)^{\infty}$ with $\bar I_{\infty}=0$ but exactly the same singlet, doublet, and triplet bulk probabilities as the Bernoulli chain so that $\bar I_3=\ln 2$; and the periodic chain $(12)^{\infty}$ with $\bar I_{\infty}=\bar I_3=0$.

\begin{figure}[h]
\centerline{\scalebox{0.45}{\includegraphics{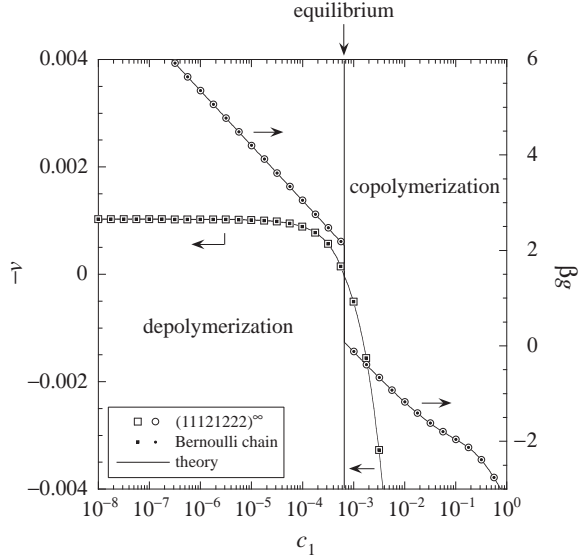}}}
\caption{Growth and depolymerization with the rate constants~(\ref{Ex1}) and the concentration $c_2=0.005$: The depolymerization velocity $-v$ (open and filled squares, vertical axis on the left-hand side) and the free enthalpy per monomeric unit rescaled by thermal energy $\beta g$ (open and filled circles, vertical axis on the right-hand side) versus the concentration $c_1$ in the cases of an initial copolymer taken as the periodic chain $(11121222)^{\infty}$ (open symbols) and a Bernoulli chain of probabilities $\bar\mu_1(1)=\bar\mu_1(2)=0.5$ (filled symbols). The dots are the numerical simulation data obtained with a sample of $10^3$ sequences of initial length $l_0=10^6$ and undergoing $2\times10^6$ reactive events.  The solid lines are the corresponding theoretical results.  The coincidence of open and filled symbols is the evidence that the Bernoulli chain and $(11121222)^{\infty}$ have precisely the same depolymerization properties, as explained in the text.}
\label{fig4}
\end{figure}

\begin{figure}[h]
\centerline{\scalebox{0.45}{\includegraphics{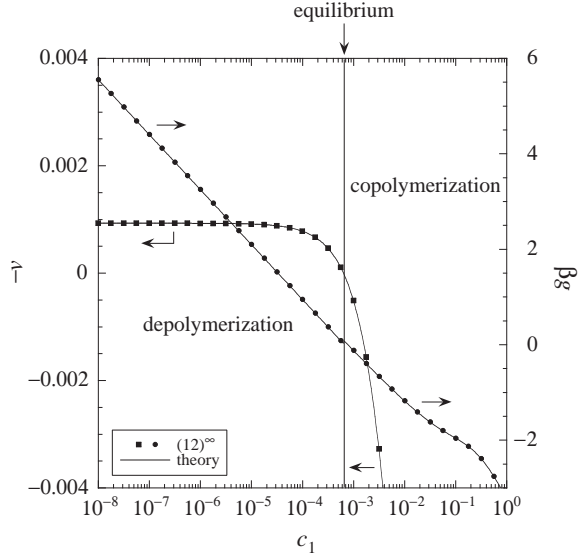}}}
\caption{Growth and depolymerization with the rate constants~(\ref{Ex1}) and the concentration $c_2=0.005$: The depolymerization velocity $-v$ (filled squares, vertical axis on the left-hand side) and the free enthalpy per monomeric unit rescaled by thermal energy $\beta g$ (filled circles, vertical axis on the right-hand side) versus the concentration $c_1$ in the case of an initial copolymer taken as the periodic chain $(12)^{\infty}$. The dots are the numerical simulation data obtained with a sample of $10^3$ sequences of initial length $l_0=10^6$ and undergoing $2\times10^6$ reactive events.  The solid lines are the corresponding theoretical results.}
\label{fig5}
\end{figure}

Figures~\ref{fig4} and~\ref{fig5} show the depolymerization velocity~$-v$ and the dimensionless free enthalpy per monomeric unit $\beta g$ as a function of the concentration $c_1$ from the growth regime down to the depolymerization regime across the critical equilibrium concentration $c_{1,{\rm eq}}\simeq 0.00064027$.   For the three chains, the velocity changes its sign across the transition in agreement with the theoretical results~(\ref{v2}) and~(\ref{v-depolym}).

In Fig.~\ref{fig4}, we see the remarkable result that the depolymerization velocity as well as the free-enthalpy per monomeric unit take precisely the same values for Bernoulli chains as for the periodic chain $(11121222)^{\infty}$ because they have the same triplet bulk probabilities $\bar\mu_3(m_1m_2m_3)=1/8$.  This is indeed expected from Eqs.~(\ref{v-depolym}) and~(\ref{eps-depolym}) with $k=2$.  We also observe that the free enthalpy per monomeric unit undergoes an important jump at the critical equilibrium concentration between the values $\beta g \leq \beta g_{\rm eq}=D_{\rm eq} \simeq 0.093$ in the growth regime and those $\beta \bar g \geq \beta \bar g_0 \simeq 2.1$ for the depolymerization of a Bernoulli chain or the periodic chain $(11121222)^{\infty}$, where the notation $\beta\bar g_0=\lim_{v\to 0} \beta \bar g$ is used.

However, the jump is much smaller in Fig.~\ref{fig5} for the depolymerization of the periodic chain $(12)^{\infty}$, in which case $\beta \bar g \geq \beta \bar g_0 \simeq 0.02$.  The reason is that the periodic chain $(12)^{\infty}$ has statistical properties $\bar\mu^{(k)}$ closer to the equilibrium probability distribution $\mu^{\rm (eq)}$, which is determined by the rates according to Eq.~(\ref{mu_eq}), than a Bernoulli chain or the periodic chain $(11121222)^{\infty}$.

In every case, the bounds~(\ref{KL-div-2}) and~(\ref{KL-div-k-3}) are satisfied in agreement with  Landauer's principle.  Indeed, $\beta \bar g \geq \beta \bar g_0 \simeq 2.1 \geq \bar I_3=\ln 2\simeq 0.693$  for Bernoulli chains and the periodic chain $(11121222)^{\infty}$, while $\beta \bar g \geq \beta \bar g_0 \simeq 0.02 \geq \bar I_3=0$ for the periodic chain $(12)^{\infty}$.

\begin{figure}[h]
\centerline{\scalebox{0.5}{\includegraphics{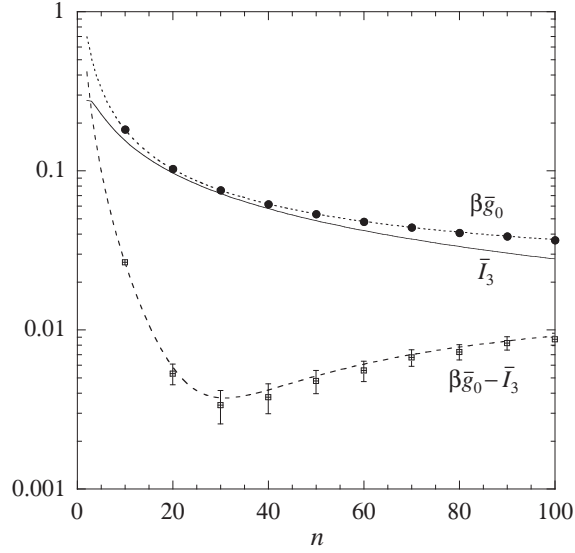}}}
\caption{Depolymerization with the rate constants~(\ref{Ex1}) and the concentration $c_2=0.005$:  The minimum free enthalpy of depolymerization rescaled by thermal energy $\beta \bar g_0$ at zero velocity $v=0$ (filled squares and dotted line) and the triplet Shannon information per monomeric unit $\bar I_3$ (solid line) for periodic sequences $[(12)^n2]^{\infty}$ versus the number $n$.  The plot also shows the difference between them (crossed squares and dashed line).  The dots are the numerical simulation data obtained with one sequence of initial length $5\times 10^6$ and undergoing $10^7$ reactive events.  The lines are the corresponding theoretical results.}
\label{fig6}
\end{figure}

In order to further test the bound~(\ref{KL-div-k-3}), Figure~\ref{fig6} shows the minimum free enthalpy per monomeric unit $\beta\bar g_0$ at vanishing depolymerization velocity $v=0$ together with the triplet Shannon information $\bar I_3$ for the depolymerization of the periodic sequences $[(12)^n2]^{\infty}$ as a function of $2\leq n \leq 100$.   At the critical equilibrium concentration $c_{1,{\rm eq}}\simeq 0.00064027$ where the velocity is vanishing $v=0$, the minimum free enthalpy of depolymerization $\beta\bar g_0$ can be calculated with Eq.~(\ref{eps-depolym}) for the periodic sequences $[(12)^n2]^{\infty}$ to be equal to
\be
\beta \bar g_0 \simeq \frac{1}{2n+1}\left( 0.04038\, n + 3.40120 \right) \simeq 0.02019 +\frac{1.69050}{n} + O(n^{-2}) \, ,
\ee
although the triplet Shannon information can be evaluated with Eq.~(\ref{I-k+1}) for $k=3$ as
\be
\bar I_3 = \frac{1}{2n+1}\left[ n \, \ln n - (n-1) \, \ln(n-1)\right] \simeq \frac{\ln({\rm e}\, n)}{2n+1}+O(n^{-2}) \, .
\ee
For the sequence $(122)^{\infty}$ at $n=1$, these quantities take the values $\beta \bar g_0\simeq 1.1472$ and $\bar I_3=0$.  For $n\to\infty$, we recover the aforementioned values $\beta\bar g_0\simeq 0.02$ and $\bar I_3=0$ of the periodic chain $(12)^{\infty}$.  

The difference $\beta\bar g_0-\bar I_3$ is also depicted in Fig.~\ref{fig6}, which is the smallest for $n=31$, but always positive.  Accordingly, the bound $\beta \bar g\geq \bar I_3$ determined by the order $k$ of the kinetic scheme is always satisfied in agreement with Eq.~(\ref{KL-div-k-3}).

\subsection{Growth of period-three copolymers and dissolution}
\label{Subsec-Ex2}

The rate constants of the second example are taken as
\be
\begin{array}{llll}
k_{+1\vert 11}=0.05 \, , &
k_{+1\vert 12}=0.1 \, , &
k_{+1\vert 21}=0.3 \, , &
k_{+1\vert 22}=2 \, , \\
k_{+2\vert 11}=0.4  \, , &
k_{+2\vert 12}=3  \, , &
k_{+2\vert 21}=4  \, , &
k_{+2\vert 22}=0.02  \, , \\
k_{-1\vert 11}=0.05 \, , &
k_{-1\vert 12}=0.02 \, , &
k_{-1\vert 21}=0.01 \, , &
k_{-1\vert 22}=0.004 \, , \\
k_{-2\vert 11}=0.03 \, , &
k_{-2\vert 12}=0.04 \, , &
k_{-2\vert 21}=0.005 \, , &
k_{-2\vert 22}=0.1 \, .
\end{array}
\label{Ex2}
\ee
Here, the concentration of monomers $2$ is fixed to the value: $c_2=0.1$, while the concentration $c_1$ is varying.

\subsubsection{Growth}

The example~(\ref{Ex2}) illustrates the growth of a copolymer of period three $(122)^{\infty}$ in the limit where the only non-vanishing rates would be $k_{\pm 1\vert 22}$, $k_{\pm 2\vert 12}$ and $k_{\pm 2\vert 21}$.  Such sequences can be synthesized by experimental techniques~\cite{SMNK10}.

\begin{figure}[h]
\centerline{\scalebox{0.5}{\includegraphics{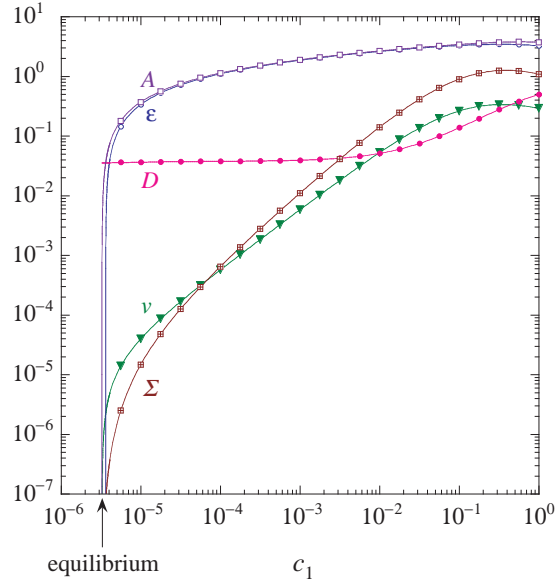}}}
\caption{Growth of a copolymer with the rate constants~(\ref{Ex2}) and the concentration $c_2=0.1$:  The mean growth velocity $v$, the free-energy driving force $\epsilon$, the Shannon disorder per monomeric unit $D$, the mean affinity $A=\epsilon+D$, and the thermodynamic entropy production $\Sigma=d_{\rm i}S/dt=A\, v$ are plotted versus the concentration $c_1$.  The dots are the numerical simulation data obtained with a sample of $10^4$ sequences grown with $10^6$ reactive events.  The solid lines are the corresponding theoretical results.}
\label{fig7}
\end{figure}

\begin{figure}[h]
\centerline{\scalebox{0.5}{\includegraphics{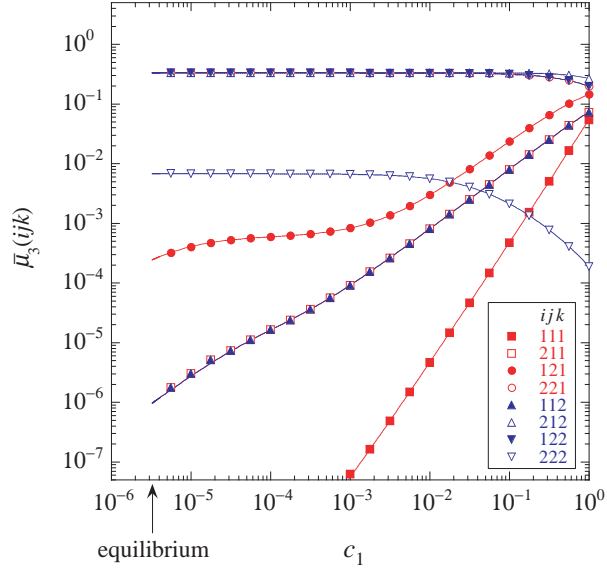}}}
\caption{Growth of a copolymer with the rate constants~(\ref{Ex2}) and the concentration $c_2=0.1$:  The bulk probabilities $\bar\mu_3(m_1m_2m_3)$ of the triplets $ijk=m_1m_2m_3$ are plotted versus the concentration $c_1$.  The dots are the numerical simulation data obtained with a sample of $10^4$ sequences grown with $10^6$ reactive events.  The solid lines are the corresponding theoretical results.}
\label{fig8}
\end{figure}

Here also, the partial velocities are efficiently obtained by iterating Eqs.~(\ref{iter2}).  They are used to get the tip and conditional probabilities of the Markov chain, allowing us to determine the mean growth velocity~(\ref{v2}), the free energy driving force~(\ref{eps}), the Shannon disorder~(\ref{D}), the mean affinity $A=\epsilon+D$, and the thermodynamic entropy production~(\ref{entrprod3}), which are depicted in Fig.~\ref{fig7} as a function of the concentration $c_1$.  Here, the critical equilibrium concentration where the mean velocity vanishes together with the mean affinity and the entropy production takes the value $c_{1,{\rm eq}}\simeq 3.2643\times 10^{-6}$ where the Shannon disorder is given by $D_{\rm eq}=-\epsilon_{\rm eq}\simeq 0.035125$.  In this example, the free-energy driving force vanishes at the concentration $c_{1,\epsilon=0}\simeq 3.6306\times 10^{-6}$.  The growth is driven by the entropic effect of sequence disorder in the interval $c_{1,{\rm eq}} < c_1 \leq c_{1,\epsilon=0}$ and by the free-energy driving force if $c_{1,\epsilon=0} < c_1$, in accordance with the predictions of Ref.~\cite{AG08}.

The triplet bulk probabilities $\bar\mu_3(m_1m_2m_3)$ obtained by Eq.~(\ref{bulk_prob}) are shown in Fig.~\ref{fig8}.  The probability is shared between the three dominant bulk probabilities $\bar\mu_3(122)\simeq \bar\mu_3(212)\simeq \bar\mu_3(221)\simeq 0.33$, as expected for copolymers of near periodicity $(122)^{\infty}$.

\begin{figure}[h]
\centerline{\scalebox{0.5}{\includegraphics{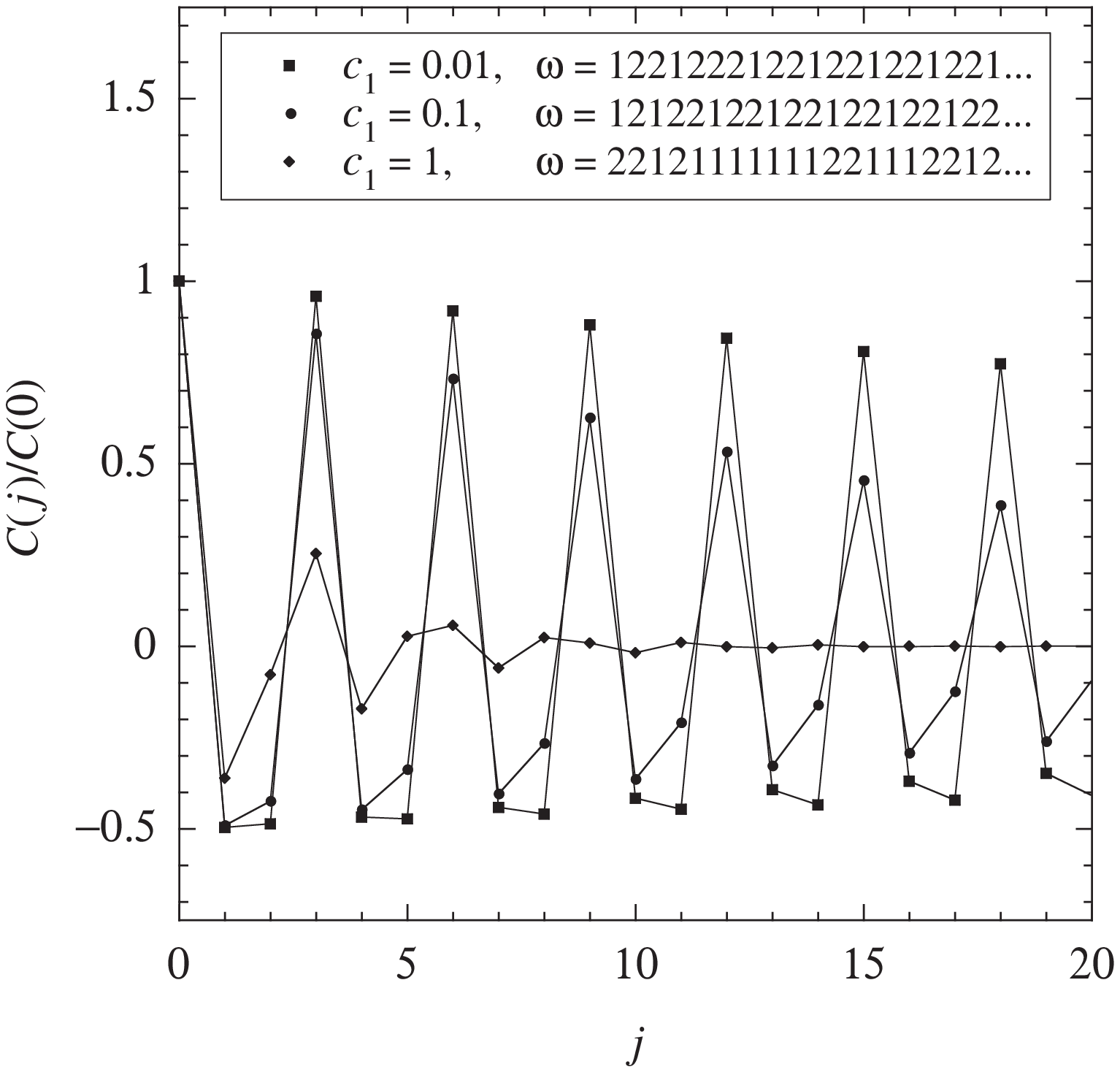}}}
\caption{Growth of a copolymer with the rate constants~(\ref{Ex2}) and the concentration $c_2=0.1$:  The normalized correlation functions~(\ref{correl-num}) versus the distance $j$ between successive monomeric units for the concentrations $c_1=0.01, 0.1,$~and~$1$. The dots are the numerical simulation data obtained with a sample of $N=10^3$ sequences of total length $L\simeq 7.6\times 10^5$ if $c_1=0.01$, $L\simeq 9.1\times 10^5$ if $c_1=0.1$, and $L\simeq 8.5\times 10^5$ if $c_1=1$.  The solid lines are the corresponding theoretical results.  The inset shows an example of the growing sequence $\omega$ for each concentration.}
\label{fig9}
\end{figure}

To confirm this conclusion, the correlation functions~(\ref{correl-num}) are depicted in Fig.~\ref{fig9} for the concentrations $c_1=0.01, 0.1$ and~$1$.  Here, we observe damped oscillations of period three.  The eigenvalues of the matrix~(\ref{M}) formed by the conditional probabilities $\mu(m_1\vert m_2m_3)$ are indeed given by
\be
\begin{array}{lllll}
&c_1 = 0.01  : & \Lambda_1=1 \, , & \Lambda_{2,3} \simeq 0.986 \, \exp(\pm i \,119.8^{\circ}) \, ,& \Lambda_4 \simeq 0.00335 \, ; \\
&c_1 = 0.1  : & \Lambda_1=1 \, ,& \Lambda_{2,3} \simeq 0.950 \, \exp(\pm i \,120.7^{\circ}) \, ,& \Lambda_4 \simeq 0.0329 \, ; \\
&c_1 = 1  : & \Lambda_1=1 \, ,& \Lambda_{2,3} \simeq 0.680 \, \exp(\pm i \,128.0^{\circ})  \, ,& \Lambda_4 \simeq 0.266 \, ;
\end{array}
\ee
which explains the behavior seen in Fig.~\ref{fig9}.

\subsubsection{Dissolution}

Depolymerization happens for concentrations below equilibrium: $c_1<c_{1,{\rm eq}}$ with $c_{1,{\rm eq}}\simeq 3.2643\times 10^{-6}$.  

\begin{figure}[h]
\centerline{\scalebox{0.45}{\includegraphics{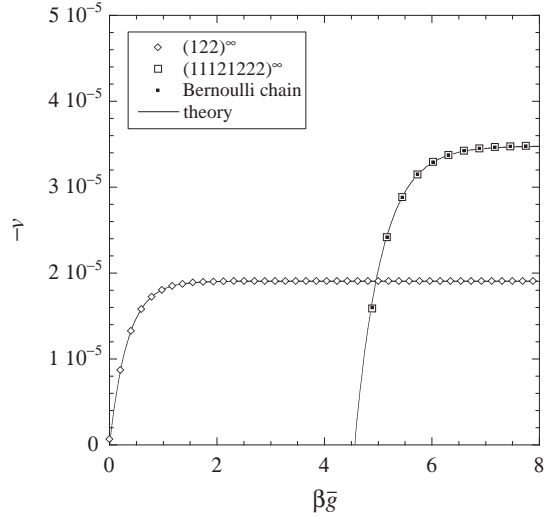}}}
\caption{Depolymerization with the rate constants~(\ref{Ex2}) and the concentration $c_2=0.1$: The depolymerization velocity $-v$ versus the free enthalpy per monomeric unit rescaled by thermal energy $\beta \bar g$ for the initial periodic chains $(122)^{\infty}$ (open diamonds) and $(11121222)^{\infty}$ (open squares), as well as for a Bernoulli chain of probabilities $\bar\mu_1(1)=\bar\mu_1(2)=0.5$ (filled squares).  The dots are the numerical simulation data obtained with a sample of $10^3$ sequences of initial length $l_0=5\times10^5$ and undergoing $10^6$ reactive events.  The solid lines are the corresponding theoretical results.}
\label{fig10}
\end{figure}

Figure~\ref{fig10} shows the depolymerization velocity~(\ref{v-depolym}) as a function of the rescaled free enthalpy~(\ref{eps-depolym}) for three different initial copolymers: a Bernoulli chain of probabilities $\bar\mu_1(1)=\bar\mu_1(2)=0.5$ with $\bar I_{\infty}=\bar I_3=\ln 2$; the periodic chain $(11121222)^{\infty}$ with $\bar I_{\infty}=0$ but exactly the same Shannon information $\bar I_3=\ln 2$ as the Bernoulli chain; and the periodic chain $(122)^{\infty}$ with $\bar I_3=\bar I_{\infty}=0$.  Since the Bernoulli chain and the periodic chain $(11121222)^{\infty}$ have the same triplet bulk probabilities $\bar\mu_3(m_1m_2m_3)=1/8$, their depolymerization velocity and free enthalpy per monomeric unit coincide, as seen in Fig.~\ref{fig10} and as predicted by Eqs.~(\ref{v-depolym}) and~(\ref{eps-depolym}).  Moreover, the bound~(\ref{KL-div-k-3}) is satisfied for the Bernoulli chain and $(11121222)^{\infty}$ because
\be
\beta\bar g_0 \simeq 4.57 \geq \bar I_3=\ln 2\simeq 0.693\, ,
\ee
where $\beta\bar g_0$ is the minimum value of the rescaled free enthalpy.
However, the gap between $\beta\bar g_0$ and $\bar I_3$ is relatively large since the Bernoulli chain and $(11121222)^{\infty}$ have triplet probabilities that are significantly different from those generated by the kinetics~(\ref{Ex2}), which is instead favoring the formation of the period-three chain $(122)^{\infty}$.  Accordingly, the free enthalpy can take much smaller values for the depolymerization of this periodic chain, as seen in Fig.~\ref{fig10}.  Nevertheless, the bound~(\ref{KL-div-k-3}) is again satisfied, because $\bar I_3$ is vanishing for $(122)^{\infty}$:
\be
\beta\bar g_0 \simeq 0.00485 \geq \bar I_3=0\, .
\ee
Therefore, the results confirm that the bound $\beta\bar g\geq \beta\bar g_0 \geq \bar I_3$ predicted by Eq.~(\ref{KL-div-k-3}) is satisfied.

\section{Conclusion}
\label{Conclusion}

In the present paper, the theory has been developed for living copolymerization processes with attachment and detachment rates depending not only on the monomer that is newly attached to or lastly detached from the molecular chain, but also on $k$ successive monomeric units behind the tip of the copolymer.  Such processes generate the growth of $k^{\rm th}$-order Markov chains.

In the regime of steady growth, the theory allows us to determine analytically the statistical properties of the growing chains in terms of partial velocities obeying self-consistent equations.  These equations only depend on the attachment and detachment rates and they can be solved by numerical iterations.  Once the partial velocities are known, the conditional probabilities of the Markov chain as well as its bulk and tip probabilities can be deduced.  The mean growth velocity is given by averaging the partial velocities over the tip probabilities.  The diffusivity of the length during growth can also be obtained.  These results are provided by extending the method already developed for the growth of Bernoulli and first-order Markov chains in Refs.~\cite{AG09,GA14}.

With the probability distribution of the monomeric sequences in the growing macromolecular chain, the nonequilibrium thermodynamics of the copolymerization process can be established.  The entropy production is explicitly shown to be given in terms of the mean growth velocity, the free energy driving force, and the Shannon disorder per monomeric unit of the $k^{\rm th}$-order Markov chain by Eq.~(\ref{entrprod3}).  Accordingly, the growth can be driven either by the free-energy driving force away from equilibrium or by the sequence disorder as equilibrium is approached \cite{AG08}.  A fluctuation relation is also deduced for the probability distribution of the random numbers of multiplets incorporated in the chain.

Necessary and sufficient conditions are formulated to find the equilibrium state where the entropy production vanishes together with the mean growth velocity.  This latter can become negative in the regime of depolymerization.  Interestingly, the properties of depolymerization are determined by the sequence of the initial macromolecular chain, which is typically different from the sequences self-generated during growth.  Indeed, the initial chain is synthesized under conditions different from those prevailing in the solution where depolymerization happens.  In this regime, analytical formulas are also obtained for the mean velocity of depolymerization and the thermodynamic entropy production, which both depend only on the statistical distribution of the multiplets composed of $k+1$ monomeric units and no more.  Consequently, the entropy production of depolymerization is always larger than the Shannon information per monomeric units in these multiplets and, thus, always larger than the overall Shannon information per monomeric unit.  Therefore, Landauer's principle is satisfied, according to which the thermodynamic entropy production of a process erasing some molecular structure cannot be smaller than the information it may contain \cite{L61}.

Theory is compared with numerical simulations by Gillespie's kinetic Monte Carlo algorithm \cite{G76,G77} for two illustrative examples of $2^{\rm nd}$-order kinetics, leading to the growth of period-two and period-three copolymers composed of two monomeric species or the corresponding depolymerization.  In both examples, theory is in nice agreement with the numerical simulations, justifying the assumptions of stationary growth or depolymerization at the basis of the analytical results.  In the examples, the different quantities of interest are studied as a function of monomeric concentrations in the surrounding solution for the regime of steady growth and depolymerization.  In between, there exists a critical value of concentration corresponding to the state of thermodynamic equilibrium.  Although the mean velocity remains continuous across the transition between growth and depolymerization, the entropy production per monomeric unit does not, its discontinuity depending on the initial copolymer sequence that is depolymerized.  This discontinuity can be reduced if the initial sequence has statistical properties close to the sequence that would naturally be grown by the kinetic process.  The study of these illustrative examples confirms that the mean depolymerization velocity as well as the free enthalpy per monomeric unit and, thus, the entropy production only depend on the statistics of triplets in the initial sequence, the length $k+1$ of these multiplets corresponding to the order $k$ of the kinetic scheme (and of the Markov chains in the growth regime).

To conclude, this work completes the analysis of free living copolymerization processes of arbitrary order started in Refs.~\cite{AG09,GA14}.  The analytical formulas are spectacularly more efficient than Monte Carlo simulations so that they can be used to quickly predict the composition of copolymers once the reaction rates of their synthesis by living copolymerization are known.  

Beyond, the present theory can also be applied to study the autonomous processes of living copolymerization with a template \cite{AG08}, or the cyclic processes used for the synthesis of sequence-controlled copolymers with encoded information \cite{LOLS13,RMLCVL15}.

\begin{acknowledgments}
The author thanks David Andrieux and Yves Geerts for useful discussions.
This research is financially supported by the Universit\'e Libre de Bruxelles, the FNRS-F.R.S., and the Belgian Federal Government under the Interuniversity Attraction Pole project P7/18 ``DYGEST".
\end{acknowledgments}

\newpage

\appendix

\section{Solving the kinetic equations}
\label{AppA}

\subsection{The growth of $1^{\rm st}$-order Markov chains}

The general method described in Section~\ref{Growth} to solve the master equations~(\ref{master}) have already been used in Ref.~\cite{GA14} to deduce Eqs.~(\ref{iter})-(\ref{cond_prob}) with $k=1$.  In this case, the attachment and detachment rates only depend on the previously incorporated monomeric units $w_{\pm m_l\vert m_{l-1}}$ and first-order Markov chains are generated.

\subsection{The growth of $2^{\rm nd}$-order Markov chains}

Remarkably, the method of Ref.~\cite{GA14} extends {\it mutatis mutandis} to higher orders $k$.  Since the equations become quite complicated, we start developing the method to the case $k=2$ where the attachment and detachment rates depend on the last two monomeric units $w_{\pm m_l\vert m_{l-1}m_{l-2}}$.  Summing the master equations~(\ref{master}) over sequences $m_1m_2\cdots m_{l-r}$ with $r=0,1,2,...$, we obtain a hierarchy of equations for the probabilities~(\ref{p0}), (\ref{p1}), (\ref{p2}),...  This hierarchy is summarized with Eq.~(\ref{Eq-P}).  Replacing with the particular solution and letting the wavenumber going to zero, the hierarchy reduces to Eq.~(\ref{Phi_0}).  This hierarchy is divided by the constant $\phi_0(\emptyset)$ to get the equations
\be
 ({\boldsymbol{\mathsf A}}+{\boldsymbol{\mathsf B}}-{\boldsymbol{\mathsf C}})\cdot\pmb{\mu}=0 
 \label{hierarchy_mu}
 \ee
 for the time-independent probabilities~(\ref{mu1}), (\ref{mu2}), (\ref{mu3}), etc.  The first few equations of this hierarchy read:
 \bea
 r=1:\qquad 0&=&\sum_{m_{l-2}m_{l-1}} w_{+m_l\vert m_{l-1}m_{l-2}}\, \mu(m_{l-2}m_{l-1}) + \sum_{m_{l-1}m_{l+1}} w_{-m_{l+1}\vert m_{l}m_{l-1}}\, \mu(m_{l-1}m_{l}m_{l+1}) \nonumber\\
 &&- \sum_{m_{l-2}m_{l-1}} w_{-m_l\vert m_{l-1}m_{l-2}}\, \mu(m_{l-2}m_{l-1}m_{l}) - \sum_{m_{l-1}m_{l+1}} w_{+m_{l+1}\vert m_{l}m_{l-1}}\, \mu(m_{l-1}m_{l}) \, , \nonumber\\
&&\label{Eq_r=1} \\
 r=2:\qquad 0&=&\sum_{m_{l-2}} w_{+m_l\vert m_{l-1}m_{l-2}}\, \mu(m_{l-2}m_{l-1}) + \sum_{m_{l+1}} w_{-m_{l+1}\vert m_{l}m_{l-1}}\, \mu(m_{l-1}m_{l}m_{l+1}) \nonumber\\
 &&- \sum_{m_{l-2}} w_{-m_l\vert m_{l-1}m_{l-2}}\, \mu(m_{l-2}m_{l-1}m_{l}) - \sum_{m_{l+1}} w_{+m_{l+1}\vert m_{l}m_{l-1}}\, \mu(m_{l-1}m_{l}) \, , \label{Eq_r=2} \\
 r=3:\qquad 0&=&w_{+m_l\vert m_{l-1}m_{l-2}}\, \mu(m_{l-2}m_{l-1}) + \sum_{m_{l+1}} w_{-m_{l+1}\vert m_{l}m_{l-1}}\, \mu(m_{l-2}m_{l-1}m_{l}m_{l+1}) \nonumber\\
 &&- \bigg(w_{-m_l\vert m_{l-1}m_{l-2}} + \sum_{m_{l+1}} w_{+m_{l+1}\vert m_{l}m_{l-1}}\bigg)\, \mu(m_{l-2}m_{l-1}m_{l}) \, , \label{Eq_r=3} \\
r=4:\qquad 0&=&w_{+m_l\vert m_{l-1}m_{l-2}}\, \mu(m_{l-3}m_{l-2}m_{l-1}) + \sum_{m_{l+1}} w_{-m_{l+1}\vert m_{l}m_{l-1}}\, \mu(m_{l-3}m_{l-2}m_{l-1}m_{l}m_{l+1}) \nonumber\\
 &&- \bigg( w_{-m_l\vert m_{l-1}m_{l-2}}+ \sum_{m_{l+1}} w_{+m_{l+1}\vert m_{l}m_{l-1}}\bigg)\, \mu(m_{l-3}m_{l-2}m_{l-1}m_{l}) \, , \label{Eq_r=4}\\
 &&\qquad\qquad\qquad\qquad\qquad\qquad\qquad\qquad\qquad\vdots \nonumber
\eea
Because of Eqs.~(\ref{mu-mu}), (\ref{mu-mu-mu}),..., summing Eq.~(\ref{Eq_r=4}) over $m_{l-3}$ gives Eq.~(\ref{Eq_r=3}), summing Eq.~(\ref{Eq_r=3}) over $m_{l-2}$ gives Eq.~(\ref{Eq_r=2}), and summing Eq.~(\ref{Eq_r=2}) over $m_{l-1}$ gives Eq.~(\ref{Eq_r=1}).  We notice that summing Eq.~(\ref{Eq_r=1}) over $m_{l}$ gives $0=0$ for $r=0$.  The key observation is that the equations for $r=3,4,...$ are all similar to each other and they can thus be solved with the assumption of factorization
\be
\mu(m_1m_2m_3\cdots m_{l-2}m_{l-1}m_l) =\mu(m_1\vert m_2 m_3) \, \mu(m_2\vert m_3 m_4) \cdots \mu(m_{l-2}\vert m_{l-1} m_{l}) \, \mu(m_{l-1}m_{l}) \, ,
\ee
as for a $2^{\rm nd}$-order Markov chain in terms of the $M^3$ conditional probabilities $\mu(m_{l-2}\vert m_{l-1} m_{l})$ and $M^2$ tip probabilities $\mu(m_{l-1}m_{l})$.  Under this assumption, all the equations $r=3,4,...$ of the hierarchy reduce to the equation for $r=3$.  Therefore, the conditional and tip probabilities should satisfy the first three equations for $r=1,2,3$, which become:
 \bea
 r=1:\qquad 0&=&\sum_{m_{l-2}m_{l-1}} w_{+m_l\vert m_{l-1}m_{l-2}}\, \mu(m_{l-2}m_{l-1}) \nonumber\\
&& + \sum_{m_{l-1}m_{l+1}} w_{-m_{l+1}\vert m_{l}m_{l-1}}\, \mu(m_{l-1}\vert m_{l}m_{l+1})\, \mu(m_{l}m_{l+1}) \nonumber\\
 &&- \sum_{m_{l-1}}\bigg[\sum_{m_{l-2}} w_{-m_l\vert m_{l-1}m_{l-2}}\, \mu(m_{l-2}\vert m_{l-1}m_{l})+ \sum_{m_{l+1}} w_{+m_{l+1}\vert m_{l}m_{l-1}}\bigg] \mu(m_{l-1}m_{l}) \, , \nonumber\\
&&\label{Eq_r=1_b} \\
 r=2:\qquad 0&=&\sum_{m_{l-2}} w_{+m_l\vert m_{l-1}m_{l-2}}\, \mu(m_{l-2}m_{l-1}) \nonumber\\
&& + \sum_{m_{l+1}} w_{-m_{l+1}\vert m_{l}m_{l-1}}\, \mu(m_{l-1}\vert m_{l}m_{l+1}) \, \mu(m_{l}m_{l+1})\nonumber\\
 &&- \bigg[\sum_{m_{l-2}} w_{-m_l\vert m_{l-1}m_{l-2}}\, \mu(m_{l-2}\vert m_{l-1}m_{l}) + \sum_{m_{l+1}} w_{+m_{l+1}\vert m_{l}m_{l-1}}\bigg] \mu(m_{l-1}m_{l}) \, , \label{Eq_r=2_b} \\
 r=3:\qquad 0&=&w_{+m_l\vert m_{l-1}m_{l-2}}\, \mu(m_{l-2}m_{l-1}) \nonumber\\
&& + \mu(m_{l-2}\vert m_{l-1}m_{l})\Bigg[\sum_{m_{l+1}} w_{-m_{l+1}\vert m_{l}m_{l-1}}\, \mu(m_{l-1}\vert m_{l}m_{l+1})\, \mu(m_{l}m_{l+1}) \nonumber\\
 &&- \bigg(w_{-m_l\vert m_{l-1}m_{l-2}}+ \sum_{m_{l+1}} w_{+m_{l+1}\vert m_{l}m_{l-1}}\bigg) \mu(m_{l-1}m_{l})\Bigg] \, . \label{Eq_r=3_b} 
\eea
We see that Eq.~(\ref{Eq_r=2_b}) should determine the tip probabilities $\mu(m_{l-1}m_{l})$ and Eq.~(\ref{Eq_r=3_b}) the conditional probabilities $\mu(m_{l-2}\vert m_{l-1}m_{l})$, if decoupling could be achieved between these quantities.  This is where the partial velocities~(\ref{part_v}) are introduced.  In the case $k=2$, they read
\be
v_{m_{l-2}m_{l-1}} \equiv \sum_{m_l} w_{+m_l\vert m_{l-1}m_{l-2}}  -  \frac{1}{\mu(m_{l-2}m_{l-1})}\sum_{m_l} w_{-m_l\vert m_{l-1}m_{l-2}} \, \mu(m_{l-2}\vert m_{l-1}m_{l}) \, \mu(m_{l-1}m_{l}) \, .
\label{part_v_k=2}
\ee
We observe that the first and third terms in the bracket of Eq.~(\ref{Eq_r=3_b}) actually form the partial velocity $v_{m_{l-1}m_{l}}$ multiplied by $\mu(m_{l-1}m_{l})$.  Hence, Eq.~(\ref{Eq_r=3_b}) gives the conditional probabilities as
\be
\mu(m_{l-2}\vert m_{l-1} m_{l}) = \frac{w_{+m_l\vert m_{l-1}m_{l-2}}\, \mu(m_{l-2}m_{l-1})}{(w_{-m_l\vert m_{l-1}m_{l-2}}+v_{m_{l-1}m_{l}})\, \mu(m_{l-1}m_{l})} \, .
\label{cond_prob_k=2}
\ee
Inserting this result into Eq.~(\ref{part_v_k=2}), we get the self-consistent equations~(\ref{iter}) with $k=2$ for the partial velocities:
\be
v_{m_{l-2} m_{l-1}} = \sum_{m_l} \frac{w_{+m_l\vert m_{l-1}m_{l-2}}\, v_{m_{l-1}m_{l}}}{w_{-m_l\vert m_{l-1}m_{l-2}}+v_{m_{l-1}m_{l}}} \, .
\label{iter_k=2}
\ee
Summing Eq.~(\ref{cond_prob_k=2}) over $m_{l-2}$ and using the normalization condition~(\ref{cond_prob_norm}), we find the equations~(\ref{tip_prob}) with $k=2$ for the tip probabilities:
\be
\mu(m_{l-1} m_{l}) = \sum_{m_{l-2}} \frac{w_{+m_l\vert m_{l-1}m_{l-2}}}{w_{-m_l\vert m_{l-1}m_{l-2}}+v_{m_{l-1}m_{l}}} \, \mu(m_{l-2}m_{l-1}) \, .
\label{tip_prob_k=2}
\ee

We can check that Eq.~(\ref{Eq_r=2_b}) transformed with Eqs.~(\ref{part_v_k=2})-(\ref{cond_prob_k=2}) also yields Eq.~(\ref{tip_prob_k=2}), while Eq.~(\ref{Eq_r=1_b}) transformed with Eq.~(\ref{cond_prob_k=2}) becomes the trivial equation $v=v$ with the mean growth velocity~(\ref{v2}) here given by
\be
v = \sum_{m_{l-2}m_{l-1}} v_{m_{l-2}m_{l-1}} \, \mu(m_{l-2}m_{l-1})= \sum_{m_{l-1}m_{l}} v_{m_{l-1}m_{l}} \, \mu(m_{l-1}m_{l}) \, .
\label{v2_k=2}
\ee
Therefore, Eqs.~(\ref{Eq_r=1_b}) and~(\ref{Eq_r=2_b}) are satisfied once Eq.~(\ref{Eq_r=3_b}) is solved by Eqs.~(\ref{cond_prob_k=2})-(\ref{tip_prob_k=2}) in terms of the partial velocities~(\ref{part_v_k=2}), which play an essential role in providing a self-consistent solution. Q.E.D.

\subsection{The growth of $k^{\rm th}$-order Markov chains}

The general case with attachment and detachment rates $w_{\pm m_l\vert m_{l-1}\cdots m_{l-k}}$, depending on $k$ previously incorporated monomeric units, can be solved with the same method.  Now, the equations of the hierarchy~(\ref{hierarchy_mu}) are similar to each other for $r=k+1,k+2,...$:
\bea
r=k+1:&&\nonumber\\
 0&=& w_{+m_l\vert m_{l-1}\cdots m_{l-k}} \, \mu(m_{l-k}\cdots m_{l-1}) 
+\sum_{m_{l+1}} w_{-m_{l+1}\vert m_{l}\cdots m_{l-k+1}} \, \mu(m_{l-k}\cdots m_{l-1}m_{l}m_{l+1}) \nonumber\\
&& -\bigg( w_{-m_{l}\vert m_{l-1}\cdots m_{l-k}} + \sum_{m_{l+1}} w_{+m_{l+1}\vert m_{l}\cdots m_{l-k+1}}\bigg)\,  \mu(m_{l-k}\cdots m_{l-1}m_{l}) \, ,
\label{Eq_r=k+1} \\
r=k+2:&& \nonumber\\
 0&=& w_{+m_l\vert m_{l-1}\cdots m_{l-k}} \, \mu(m_{l-k-1}m_{l-k}\cdots m_{l-1}) \nonumber\\
&& +\sum_{m_{l+1}} w_{-m_{l+1}\vert m_{l}\cdots m_{l-k+1}} \, \mu(m_{l-k-1}m_{l-k}\cdots m_{l-1}m_{l}m_{l+1}) \nonumber\\
&& -\bigg( w_{-m_{l}\vert m_{l-1}\cdots m_{l-k}} + \sum_{m_{l+1}} w_{+m_{l+1}\vert m_{l}\cdots m_{l-k+1}}\bigg)\,  \mu(m_{l-k-1}m_{l-k}\cdots m_{l-1}m_{l}) \, ,
\label{Eq_r=k+2} \\
 &&\qquad\qquad\qquad\qquad\qquad\qquad\qquad\qquad\qquad\vdots \nonumber
\eea
so that they are solved with the assumption~(\ref{Markov-factor}) that the Markov chain is of $k^{\rm th}$ order.  The equation~(\ref{Eq_r=k+1}) for $r=k+1$ becomes 
\bea
&& r=k+1:\nonumber\\
 0&=& w_{+m_l\vert m_{l-1}\cdots m_{l-k}} \, \mu(m_{l-k}\cdots m_{l-1}) \nonumber\\
&&+\mu(m_{l-k}\vert m_{l-k+1}\cdots m_{l})  \, \Bigg[\sum_{m_{l+1}} w_{-m_{l+1}\vert m_{l}\cdots m_{l-k+1}} \, \mu(m_{l-k+1}\vert m_{l-k+2} \cdots m_{l+1})\, \mu(m_{l-k+2}\cdots m_{l+1}) \nonumber\\
&& -\bigg( w_{-m_{l}\vert m_{l-1}\cdots m_{l-k}} + \sum_{m_{l+1}} w_{+m_{l+1}\vert m_{l}\cdots m_{l-k+1}}\bigg)\,  \mu(m_{l-k+1}\cdots m_{l}) \Bigg] \, .
\label{Eq_r=k+1_b}
\eea
Thanks to the introduction of the partial velocities~(\ref{part_v}), Eq.~(\ref{Eq_r=k+1_b}) gives the expression~(\ref{cond_prob}) for the conditional probabilities $\mu(m_{l-k}\vert m_{l-k+1}\cdots m_{l})$.  On the one hand, the self-consistent equations~(\ref{iter}) for the partial velocities are obtained by inserting Eq.~(\ref{cond_prob}) into the definition~(\ref{part_v}) for the partial velocities.  On the other hand, 
Eq.~(\ref{tip_prob}) for the tip probabilities are given by summing Eq.~(\ref{cond_prob}) over $m_{l-k}$ and using the normalization conditions~(\ref{cond_prob_norm}).  

Here also, we can check that the equations for $r=1,2,...,k$ in the hierarchy are satisfied with the assumption~(\ref{Markov-factor}) and Eqs.~(\ref{part_v})-(\ref{cond_prob}).  Q.E.D.


\section{Proof of the fluctuation relation}
\label{AppB}

In order to prove the fluctuation relation~(\ref{FT}), we should start from the master equation
\bea
\frac{d}{dt}\, P_t(m_1\cdots m_l,\Delta{\bf N}) &=& w_{+m_l\vert m_{l-1}\cdots m_{l-k}} \, P_t(m_1\cdots m_{l-1},\Delta{\bf N}-{\bf 1}_{m_{l-k}\cdots m_l}) \nonumber\\
&&+\sum_{m_{l+1}=1}^M w_{-m_{l+1}\vert m_{l}\cdots m_{l-k+1}} \, P_t(m_1\cdots m_{l+1},\Delta{\bf N}+{\bf 1}_{m_{l-k+1}\cdots m_{l+1}}) \nonumber\\
&&- \left( w_{-m_{l}\vert m_{l-1}\cdots m_{l-k}} + \sum_{m_{l+1}=1}^M w_{+m_{l+1}\vert m_{l}\cdots m_{l-k+1}}\right)  P_t(m_1\cdots m_{l},\Delta{\bf N}) \, , \nonumber\\
&&
\label{master-N}
\eea
for the probability $P_t(m_1\cdots m_l,\Delta{\bf N})$ that the chain has the sequence $m_1\cdots m_l$ with $l\gg k$ at the time $t$ and that the $M^{k+1}$ counters of the different possible multiplets $m_j\cdots m_{j+k}$ of length $k+1$ take the values $\Delta{\bf N} = \left\{ \Delta N_{m_j\cdots m_{j+k}}\right\}$.  The notation
\be
(\Delta{\bf N}\pm{\bf 1}_{m_{i}\cdots m_{i+k}})_{m_j\cdots m_{j+k}}= 
\left\{
\begin{array}{ll}
\Delta N_{m_j\cdots m_{j+k}}\, , &\mbox{if} \quad m_j\cdots m_{j+k}\neq m_{i}\cdots m_{i+k} \\
\Delta N_{m_j\cdots m_{j+k}}\pm 1\, , &\mbox{if} \quad m_j\cdots m_{j+k}=m_{i}\cdots m_{i+k}
\end{array}
\right. 
\ee
is used in Eq.~(\ref{master-N}).

On the one hand, we notice that the master equation~(\ref{master}) is recovered from Eq.~(\ref{master-N}) by summing over $\Delta{\bf N}$.  On the other hand, the probability distribution appearing in the fluctuation relation~(\ref{FT}) is defined as
\be
p_t(\Delta{\bf N}) \equiv \sum_{m_1\cdots m_l} P_t(m_1\cdots m_l,\Delta{\bf N}) \, .
\label{ptN}
\ee
In the long-time limit, the same factorization as in Eq.~(\ref{p-mu_factor}) is expected:
\be
P_t(m_1\cdots m_{l-1}m_l,\Delta{\bf N})\simeq p_t(\Delta{\bf N}) \, \mu(m_1 \cdots m_{l-1} m_l) \, .
\label{M.fact}
\ee
Therefore, the probability distribution~(\ref{ptN}) evolves in time according to
\be
\frac{dp_t}{dt} = \hat L \, p_t
\ee
with the linear operator
\bea
\hat L &\equiv& \sum_{m_{l-k}\cdots m_l} \Big[ w_{+m_l\vert m_{l-1}\cdots m_{l-k}} \, \mu(m_{l-k}\cdots m_{l-1}) \, \left( \hat E^{-}_{m_{l-k}\cdots m_l}-1\right) \nonumber\\
&& + w_{-m_l\vert m_{l-1}\cdots m_{l-k}} \, \mu(m_{l-k}\vert m_{l-k+1}\cdots m_l) \, \mu(m_{l-k+1}\cdots m_l) \, \left( \hat E^{+}_{m_{l-k}\cdots m_l}-1\right) \Big]
\eea
where
\be
\hat E^{\pm}_{m_{l-k}\cdots m_l} f(\Delta{\bf N}) = f(\Delta{\bf N}\pm{\bf 1}_{m_{l-k}\cdots m_l}) \, .
\ee

The cumulant generating function of the counters $\Delta{\bf N}$ is defined as
\be
Q(\pmb{\lambda}) \equiv \lim_{t\to\infty} -\frac{1}{t} \ln \left\langle {\rm e}^{-\pmb{\lambda}\cdot\Delta{\bf N}}\right\rangle_t
\ee
in terms of the counting parameters $\pmb{\lambda}=\{\lambda_{m_{l-k}\cdots m_l}\}\in{\mathbb R}^{M^{k+1}}$.  The cumulant generating function can be obtained as the leading eigenvalue, $\hat L_{\pmb{\lambda}}\,\varphi = -Q(\pmb{\lambda})\,\varphi$, for the modified operator
\be
\hat L_{\pmb{\lambda}} \equiv {\rm e}^{-\pmb{\lambda}\cdot\Delta{\bf N}} \hat L \, {\rm e}^{+\pmb{\lambda}\cdot\Delta{\bf N}} \, ,
\ee
giving
\bea
Q(\pmb{\lambda}) &=& \sum_{m_{l-k}\cdots m_l} \Big[ w_{+m_l\vert m_{l-1}\cdots m_{l-k}} \, \mu(m_{l-k}\cdots m_{l-1}) \, \left( 1-{\rm e}^{-\lambda_{m_{l-k}\cdots m_l}}\right) \nonumber\\
&& + w_{-m_l\vert m_{l-1}\cdots m_{l-k}} \, \mu(m_{l-k}\vert m_{l-k+1}\cdots m_l) \, \mu(m_{l-k+1}\cdots m_l) \, \left( 1-{\rm e}^{+\lambda_{m_{l-k}\cdots m_l}}\right) \Big] \, .
\label{Q}
\eea
Since the sum extends over all the multiplets of length $k+1$, the indices can be arbitrarily changed as $m_{l-k}\cdots m_l \to m_j\cdots m_{j+k}$ for any integer $j$.

The net rates~(\ref{J}) or~(\ref{J2}) are obtained by differentiating the cumulant generating function~(\ref{Q}) with respect to the counting parameters:
\be
{\bf J} = \frac{\partial Q}{\partial\pmb{\lambda}}\Big\vert_{\pmb{\lambda}=0} = \lim_{t\to\infty} \frac{1}{t} \langle\Delta{\bf N}\rangle_t \, .
\ee

The remarkable property is that the cumulant generating function~(\ref{Q}) obeys the symmetry relation
\be
Q(\pmb{\lambda}) =Q({\bf A}-\pmb{\lambda}) \, ,
\label{FT.Q}
\ee
with respect to the affinities~(\ref{affinities}).  Using large-deviation theory~\cite{E85,T09}, the fluctuation relation~(\ref{FT}) is thus established. Q.E.D.


\end{document}